\documentclass[amsmath,amssymb,nofootinbib,notitlepage,superscriptaddress,twocolumn]{revtex4-1}
\pdfoutput=1

\usepackage{aas_macros,bm,graphicx,xcolor}
\usepackage[bottom]{footmisc}
\usepackage[colorlinks=true]{hyperref}
\let\originalleft\left
\let\originalright\right
\renewcommand{\left}{\mathopen{}\mathclose\bgroup\originalleft}
\renewcommand{\right}{\aftergroup\egroup\originalright}

\newcommand{\ab}[1]{\left|#1\right|}
\newcommand{\br}[1]{\left[#1\right]}
\newcommand{\cu}[1]{\left\{#1\right\}}
\newcommand{\pa}[1]{\left(#1\right)}

\newcommand{\ed}{\mathop{}\!\mathrm{d}}
\renewcommand{\O}[1]{\mathcal{O}\pa{#1}}

\definecolor{red}{RGB}{228,26,28}
\definecolor{green}{RGB}{77,175,74}
\definecolor{blue}{RGB}{55,126,184}
\definecolor{purple}{RGB}{152,78,163}
\definecolor{orange}{RGB}{255,127,0}
\definecolor{brown}{RGB}{166,86,40}
\definecolor{pink}{RGB}{247,129,191}

\begin{document}

\title{On the Observable Shape of Black Hole Photon Rings}

\author{Samuel E. Gralla}
\email{sgralla@email.arizona.edu}
\affiliation{Department of Physics, University of Arizona, Tucson, AZ 85721, USA}
\author{Alexandru Lupsasca}
\email{lupsasca@princeton.edu}
\affiliation{Princeton Gravity Initiative, Princeton University, Princeton, NJ 08544, USA}
\affiliation{Society of Fellows, Harvard University, Cambridge, MA 02138, USA}

\begin{abstract}
Motivated by the prospect of measuring a black hole photon ring, in previous work we explored the interferometric signature produced by a bright, narrow curve in the sky.  Interferometric observations of such a curve measure its ``projected position function'' $\bm{r}\cdot\bm{\hat{n}}$, where $\bm{r}$ parameterizes the curve and $\bm{\hat{n}}$ denotes its unit normal vector.  In this paper, we show by explicit construction that a curve can be fully reconstructed from its projected position, completing the argument that space interferometry can in principle determine the detailed photon ring shape.  In practice, near-term observations may be limited to the visibility \textit{amplitude} alone, which contains incomplete shape information: for convex curves, the amplitude only encodes the set of projected diameters (or ``widths'') of the shape.  We explore the freedom in reconstructing a convex curve from its widths, giving insight into the shape information probed by technically plausible future astronomical measurements.  Finally, we consider the Kerr ``critical curve'' in this framework and present some new results on its shape.  We analytically show that the critical curve is an ellipse at small spin or inclination, while at extremal spin it becomes the convex hull of a Cartesian oval.  We find a simple oval shape, the ``phoval'', which reproduces the critical curve with high fidelity over the whole parameter range.
\end{abstract}

\maketitle

\section{Introduction}

General relativity predicts that gravitational lensing will generically produce narrow ``photon rings'' on images of sources near a black hole \cite{Bardeen1973,Luminet1979,Beckwith2005,GrallaHolz2019,Johnson2020,GrallaLupsasca2020a}.  Recently, horizon-scale emission from the black hole in the galaxy M87 was resolved using ground-based interferometry \cite{EHT2019a,EHT2019b,EHT2019c,EHT2019d,EHT2019e,EHT2019f}, and future space missions may be able to measure the much narrower photon ring via its universal signature on long baselines \cite{Johnson2020,Gralla2020}.  A precise measurement of the photon ring shape could be used to infer black hole parameters and/or as a test of general relativity.

The observable signature of this shape in interferometry is the ``projected position function'' of the underlying curve \cite{Gralla2020}.  By counting of (functional) degrees of freedom, one expects to be able to reconstruct the full curve from this information, but no proof was given in Ref.~\cite{Gralla2020}.  This paper provides a simple constructive proof of this fact, completing the argument that the universal signature discussed in Refs.~\cite{Johnson2020,Gralla2020} encodes the photon ring shape and giving a straightforward method to recover it.

Measuring the \textit{full} projected position function of a photon ring requires absolute phase tracking, which may be experimentally challenging with present technology.  A more plausible near-term goal is the measurement of the visibility amplitude alone, which provides only partial information: for a closed, convex curve, the amplitude encodes only the set of projected diameters, or \textit{widths}, of the shape \cite{Gralla2020}.  The second purpose of this paper is to explore the information contained solely in the widths of a closed, convex shape.  We show that the freedom in reconstructing a closed, convex shape from its widths consists of one choice of an odd function on the circle, and we give a constructive method for producing all shapes that share a given set of widths.

The shape of astrophysical photon rings is expected to closely approximate that of the theoretical ``critical curve'' predicted by general relativity for Kerr black holes.  The third purpose of this paper is to explore the detailed shape of this critical curve as encoded in its interferometric observables.  We analytically show that the critical curve is an ellipse to quadratic order in (either) the black hole spin or observer inclination, and derive the associated ellipse parameters.  At extremal spin (and any inclination), we show that the critical curve is the convex hull of a Cartesian oval, a classical shape first studied by Descartes in 1637.  We also compute analytically the projected position function of the equatorial critical curve at any spin.  Finally, we find approximate expressions for the critical curve that provide an excellent fit over the entire parameter range, including the extremal limit.  These expressions are obtained by using a certain notion of ``shape addition'' that naturally arises in our study.  By adding together a circle, an ellipse, and a certain cuspy triangle, we obtain a closed curve that we dub the \textit{phoval}.  This four-parameter family of shapes encompasses the three-parameter critical curve of the Kerr black hole to very high accuracy.

This paper is organized as follows.  In Sec.~\ref{sec:GeneralCurves}, we show how to reconstruct a general plane curve from its universal signature on long interferometric baselines.  We then specialize in Sec.~\ref{sec:ConvexCurves} to closed, convex curves (including a certain cuspy generalization).  Next, we present a series of examples in Sec.~\ref{sec:Examples}, before finally turning to the study of the Kerr critical curve in Sec.~\ref{sec:CriticalCurve}.  In the appendices, we also connect with the Cauchy surface area theorem, review the method of implicitization of plane curves, and study the algebraic geometry of the Kerr critical curve.

\section{General curves}
\label{sec:GeneralCurves}

Consider a smooth (possibly disjoint) plane curve, together with a set of Cartesian axes.  Fix a polar angle $\varphi$ and let $\bm{\hat{u}}$ be the outward unit vector in that direction,
\begin{align}
	\label{eq:RadialVector}
	\bm{\hat{u}}=\pa{\cos{\varphi},\sin{\varphi}}.
\end{align}
This vector is not to be viewed as a vector field, but rather as a fixed vector (for each choice of $\varphi$) that can be placed at any point in the plane.

For each $\varphi\in[0,\pi)$, consider the set of points $\bm{r}_I$ on the curve such that $\bm{\hat{u}}$ is normal to the curve at these points.  Consider the generic case in which the curvature is non-vanishing at these points.\footnote{Isolated points where the curvature vanishes (and $\bm{\hat{n}}_I$ is ill-defined) are to be treated as a limit, where $\bm{\hat{n}}_I$ can change sign and the number of different points $I$ can change.  A finite segment of the curve with vanishing curvature (straight line) can be approximated by a very small curvature over the region.}  Each point has a sign $S_I$ defined by whether $\bm{\hat{u}}$ points toward ($+$) or away ($-$) from the center of curvature.  Letting $\bm{\hat{n}}_I$ denote the unit normal pointing toward the center of curvature, we have
\begin{align}
	\label{eq:Sign}
	S_I(\varphi)=\bm{\hat{n}}_I(\varphi)\cdot\bm{\hat{u}}(\varphi)
	\in\cu{\pm1}.
\end{align}
For each point, we further define the \textit{projected position},
\begin{align}
	\label{eq:ProjectedPosition}
	z_I(\varphi)=\bm{r}_I(\varphi)\cdot\bm{\hat{u}}(\varphi).
\end{align}
The information directly available in the universal interferometric signature of a narrow curve is the set $\cu{S_I(\varphi),z_I(\varphi)}$ for all $0\leq\varphi<\pi$ (see Eq.~(36) and surrounding discussion in Ref.~\cite{Gralla2020}).

Above, we considered fixing $\varphi$ and finding a list of points $\bm{r}_I$.  We may instead imagine fixing $I$ and varying $\varphi$, such that the functions $\bm{r}_I(\varphi)$ parameterize the curve.  Each segment $I$ may have a different range of $\varphi$. The direction of increasing $\varphi$ defines an orientation for each segment, but this does not necessarily provide a consistent orientation across segments that join.  In particular, the orientation flips when curves join at inflection points.  The normal $\bm{\hat{n}}_I$ is continuous over each segment $I$, and points in the direction of the center of curvature, which is given by a $90^\circ$ counter-clockwise rotation of the tangent vector $\bm{r}_I'(\varphi)$.  A Frenet-Serret formula relates the two directly by
\begin{align}
	\label{eq:FrenetSerret}
	\bm{r}_I'(\varphi)=-\mathcal{R}_I(\varphi)\bm{\hat{n}}_I'(\varphi),
\end{align}
where $\mathcal{R}_I(\varphi)>0$ is the radius of curvature at $\bm{r}_I(\varphi)$.

From Eqs.~\eqref{eq:RadialVector}, \eqref{eq:Sign} and \eqref{eq:FrenetSerret}, we obtain expressions for the tangent and normal to the curve at $\bm{r}_I(\varphi)$,
\begin{align}
	\label{eq:TangentVector}
	\bm{r}_I'(\varphi)&=S_I\mathcal{R}_I(\varphi)\pa{\sin{\varphi},-\cos{\varphi}},\\
	\label{eq:NormalVector}
	\bm{\hat{n}}_I(\varphi)&=S_I\pa{\cos{\varphi},\sin{\varphi}}.
\end{align}
Here, we dropped the $\varphi$-dependence of $S_I$, imagining that we restrict to a single segment $I$, over which it is constant.

Together, Eqs.~\eqref{eq:ProjectedPosition}, \eqref{eq:TangentVector}, and \eqref{eq:NormalVector} imply that the projected position $z_I$ and its first derivative are given by
\begin{align}
	z_I(\varphi)&=x_I(\varphi)\cos{\varphi}+y_I(\varphi)\sin{\varphi},\\
	z_I'(\varphi)&=-x_I(\varphi)\sin{\varphi}+y_I(\varphi)\cos{\varphi},
\end{align}
where $x_I$ and $y_I$ are the Cartesian positions $\bm{r}_I=(x_I,y_I)$.  Inverting gives the parameterization as
\begin{subequations}
\label{eq:NormalParameterization}
\begin{align}
	x_I(\varphi)&=z_I(\varphi)\cos{\varphi}-z_I'(\varphi)\sin{\varphi},\\
	y_I(\varphi)&=z_I(\varphi)\sin{\varphi}+z_I'(\varphi)\cos{\varphi}.
\end{align}
\end{subequations}
Taking a derivative and comparing with Eq.~\eqref{eq:TangentVector} gives the curvature radius as
\begin{align}
	\label{eq:CurvatureRadius}
	\mathcal{R}_I(\varphi)=-S_I\br{z_I(\varphi)+z_I''(\varphi)}.
\end{align}
Since Eqs.~\eqref{eq:NormalParameterization} parameterize the curve over some collection of segments $I$, this completes the proof that the entire curve can be uniquely reconstructed from the observational data.  Note that the sign $S_I$ contains redundant information; the full curve follows entirely from $z_I(\varphi)$.

\section{Closed, convex curves}
\label{sec:ConvexCurves}

\begin{figure}
	\centering
	\includegraphics[width=\columnwidth]{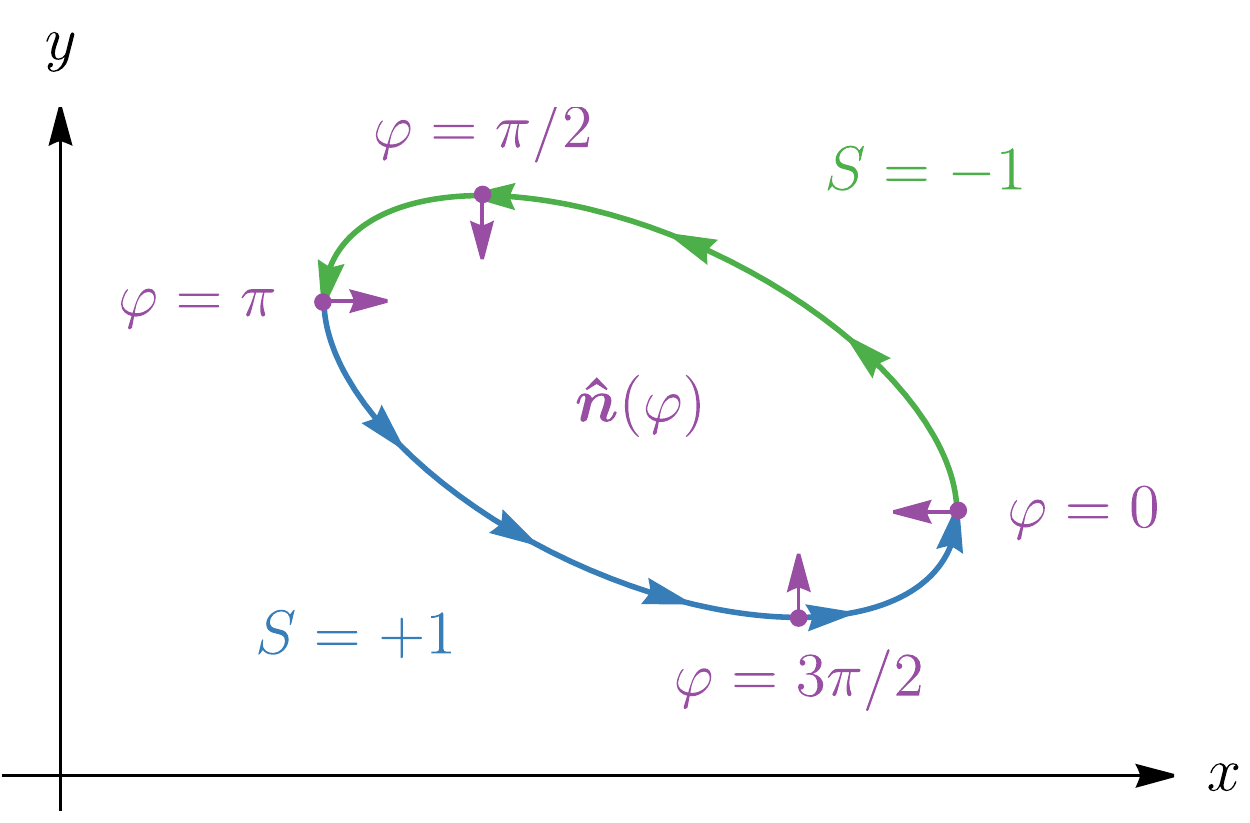}
	\caption{In the general framework capable of handling arbitrary plane curves (Sec.~\ref{sec:GeneralCurves}), a closed, convex curve is described by two segments: its ``top'' ($S=-1$, green), and its ``bottom'' ($S=+1$, blue).  Each of these segments is separately parameterized by $\varphi\in[0,\pi)$ (not shown).  In the framework of Sec.~\ref{sec:ConvexCurves}, we instead parameterize the whole curve with a single angle $\varphi$ (purple) ranging over $[0,2\pi)$.  This provides a counter-clockwise orientation around the curve (depicted with arrows).  We use a single inward-pointing normal vector $\bm{\hat{n}}$.}
	\label{fig:ConvexCurve}
\end{figure}

A convex curve is one for which every tangent line intersects the curve precisely once.  In the formalism developed above, a closed, convex curve consists of a ``top'' segment with $S_I=-1$ joined with a ``bottom'' segment with $S_I=+1$.  The angle $\varphi$ takes the full range $[0,\pi)$ on each segment, jumping discontinuously between $0$ and $\pi$ at the joining points.  We may eliminate these jumps as well as the signs by instead parameterizing the bottom segment with $\varphi\in[\pi,2\pi)$, so that the whole curve is described by $\varphi\in[0,2\pi)$---see Fig.~\ref{fig:ConvexCurve}.  That is, if an interferometric observation reveals two segments with opposite signs that smoothly join (i.e., a closed, convex curve), then it is convenient to promote $\varphi$ to the full range $[0,2\pi)$ and thereby deal with a single parameterized curve $\bm{r}(\varphi)$.

We will call this $\varphi$ the \textit{normal angle}.  It is the angle that the \textit{inward} normal vector makes with the \textit{negative} $x$-axis.  (With these conventions, $\varphi$ reduces to the usual polar angle when the curve is an origin-centered circle.)  The formulas for the tangent \eqref{eq:TangentVector} and normal \eqref{eq:NormalVector} become
\begin{align}
	\label{eq:ConvexTangent}
	\bm{r}'(\varphi)&=\mathcal{R}(\varphi)\pa{-\sin{\varphi},\cos{\varphi}},\\
	\label{eq:ConvexNormal}
	\bm{\hat{n}}(\varphi)&=-\pa{\cos{\varphi},\sin{\varphi}}.
\end{align}
These are the counter-clockwise tangent (positive orientation) and the inward normal.  One may regard Eq.~\eqref{eq:ConvexNormal} as the definition of the normal angle $\varphi$ of a convex curve (stipulating that $\bm{\hat{n}}$ is the inward unit normal).

To represent projected position for a closed, convex curve we use $f=-Sz$ instead of $z$, or equivalently,
\begin{align}
	f(\varphi)=-\bm{r}(\varphi)\cdot\bm{\hat{n}}(\varphi).
\end{align}
More explicitly, we have
\begin{align}
	\label{eq:ExplicitPosition}
	f(\varphi)=x(\varphi)\cos{\varphi}+y(\varphi)\sin{\varphi}.
\end{align}
The curvature radius \eqref{eq:CurvatureRadius} is now given by
\begin{align}
	\label{eq:ConvexCurvature}
	\mathcal{R}(\varphi)=f(\varphi)+f''(\varphi),
\end{align}
while the parameterization \eqref{eq:NormalParameterization} takes an identical form,
\begin{subequations}
\label{eq:ConvexParameterization}
\begin{align}
	x(\varphi)&=f(\varphi)\cos{\varphi}-f'(\varphi)\sin{\varphi},\\
	y(\varphi)&=f(\varphi)\sin{\varphi}+f'(\varphi)\cos{\varphi}.
\end{align}
\end{subequations}
This is just the vector $(f(\varphi),f'(\varphi))$ rotated by $\varphi$. 

The function $f(\varphi)$ contains information both about the ``curve itself'' and about the choice of Cartesian axes.  Rotations of the plane induce rotations on the circle, i.e., translations in $\varphi$.  By contrast, translations of the plane $\bm{r}\to\bm{r}+(X,Y)$ change $f(\varphi)$ by [see Eq.~\eqref{eq:ExplicitPosition}]
\begin{align}
	f\to f+X\cos{\varphi}+Y\sin{\varphi}.
\end{align}

In practice, one will typically want to compute the projected position function $f(\varphi)$ from a plane curve given in parametric form $(x(\sigma),y(\sigma))$.  The angle $\varphi$ that we have defined satisfies 
\begin{align}
	\label{eq:ParameterizedAngle}
	\tan{\varphi(\sigma)}=-\frac{x'(\sigma)}{y'(\sigma)},
\end{align}
which allows one to directly compute $\varphi(\sigma)$.  Plugging into Eq.~\eqref{eq:ExplicitPosition} then provides the projected position in terms of $\sigma$, but comparison with observations (and/or use of the framework of this section) requires knowledge of $f$ in terms of $\varphi$.  The needed inversion to find $\sigma(\varphi)$ is in general highly nontrivial.  Note, however, that one can easily plot $f(\varphi)$ as a parametric curve $(\varphi(\sigma),f(\varphi(\sigma))$.

\subsection{Extension to cuspy curves}

\begin{figure}
	\centering
	\includegraphics[width=\columnwidth]{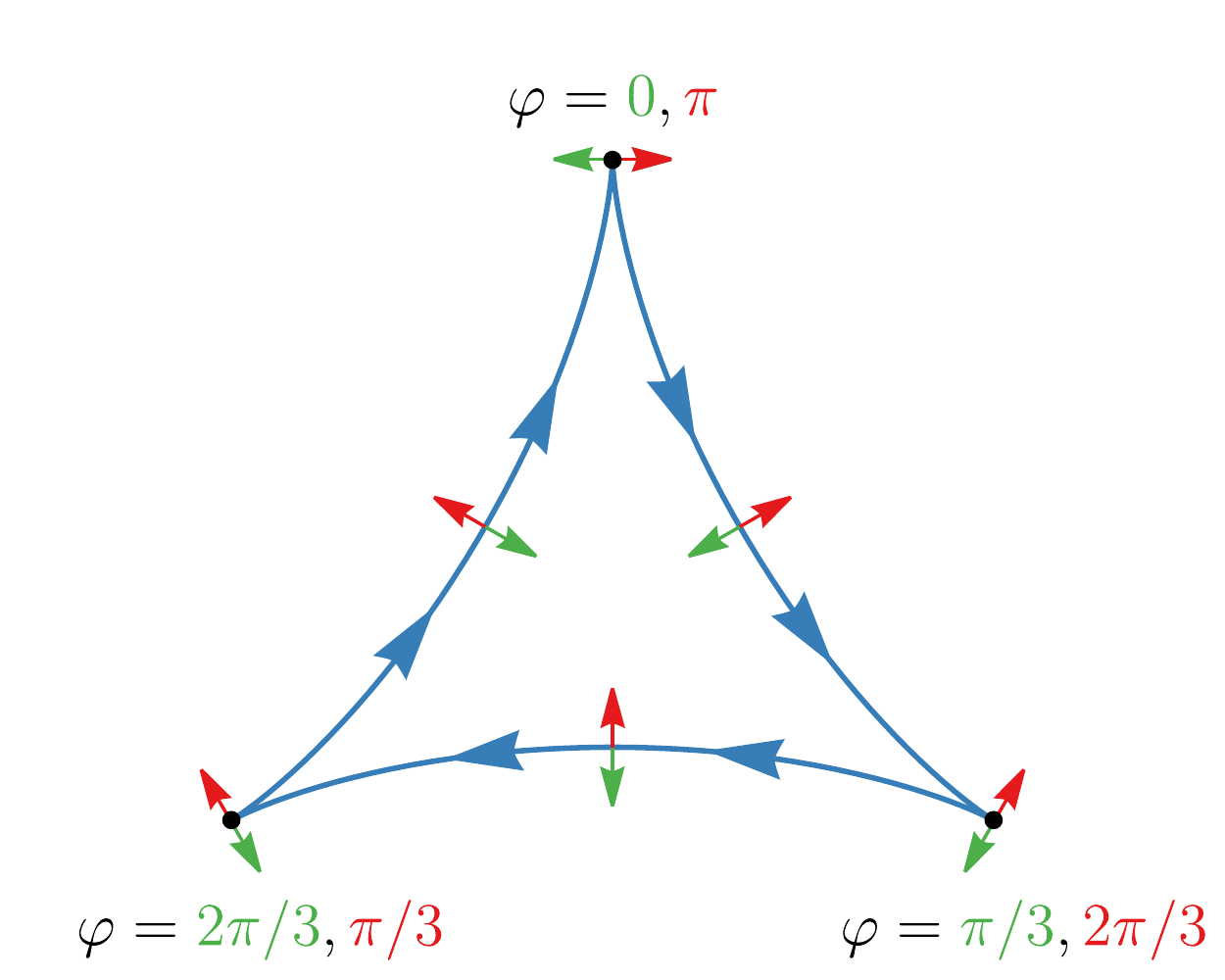}
	\caption{The framework developed for closed, convex curves is naturally extended to a class of cuspy curves by allowing $f(\varphi)$ to be an arbitrary $C^1$ function on the circle.  The parameterization \eqref{eq:ConvexParameterization} provides a consistent orientation around the curve (depicted with arrows), and the normal vector \eqref{eq:ConvexNormal} varies continuously.  In this example of $f(\varphi)=\sin{3\varphi}$, the full range $\varphi\in[0,2\pi)$ traces over the closed curve twice.  After one loop around the figure (green), the normal vector has switched sign (top point); after the second loop (red), it has returned to its original leftward orientation.}
	\label{fig:CuspyCurve}
\end{figure}

In the previous section, we showed that for a smooth convex curve, the projected position function $f$ provides a parameterization of the curve by the normal angle $\varphi$.  Conversely, any $C^1$ function $f(\varphi)$ on the circle defines a closed curve via Eqs.~\eqref{eq:ConvexParameterization}.  However, this curve is in general not convex or even simple or smooth.  To guarantee a closed, convex curve one must choose $f(\varphi)$ to be a $C^2$ function such that $f(\varphi)+f''(\varphi)$ is strictly positive.\footnote{One can also obtain a convex shape by choosing $f(\varphi)$ such that $f(\varphi)+f''(\varphi)$ is strictly negative.  However, in this case the normal vector $\bm{\hat{n}}$ defined by \eqref{eq:ConvexNormal} points outward, such that the shape is not properly parameterized by the normal angle.}  This ensures that the curvature never vanishes, proving that the closed curve is convex.

However, it is interesting (and important for the decomposition discussed below) to ask what sort of non-convex curves may be represented within this framework.  That is, suppose one chooses some $f(\varphi)$ that is $C^2$ on the circle (so as to keep the curvature radius finite), but otherwise arbitrary.  What kind of curve is defined by the parameterization \eqref{eq:ConvexParameterization}?  To examine this question, we will take Eq.~\eqref{eq:ConvexCurvature} to be the \textit{definition} of $\mathcal{R}$ (which can now be negative), and we will take Eq.~\eqref{eq:ConvexNormal} to be the \textit{definition} of $\bm{\hat{n}}$ (which can now also point away from the center of curvature).  The other equations in Sec.~\ref{sec:ConvexCurves} hold without modification.

Computing the signed curvature $k$ from Eqs.~\eqref{eq:ConvexParameterization}, we find that $k=1/|\mathcal{R}|$.  That is, the curvature is always positive, although it diverges wherever $\mathcal{R}$ vanishes, signaling the presence of a sharp cusp.  At such a cusp, the unit tangent $\bm{r}'(\varphi)/\ab{\bm{r}'(\varphi)}$ flips sign [see Eq.~\eqref{eq:ConvexTangent}].  This means that the parameter $\varphi$ provides a consistent orientation of the whole closed curve, with the tangent changing direction at cusps.  Notice, however, that the normal defined by Eq.~\eqref{eq:ConvexNormal} does not change sign: it is by definition continuous on the whole curve (but no longer always points toward the center of curvature).

In summary, the curves defined using Eqs.~\eqref{eq:ConvexParameterization} with a generic function $f(\varphi)$ that is $C^2$ on the circle are either convex, or composed of individually convex segments joined at sharp cusps. An example of the latter is shown in Fig.~\ref{fig:CuspyCurve}.

As $f(\varphi)$ is defined on the circle, it is tempting to analyze its Fourier series.  At first, this approach works beautifully: the zeroth moment is related to the perimeter of the curve (App.~\ref{app:Perimeter}), and the first moment can always be set to zero by a translation of the Cartesian origin.  However, the higher moments are not as convenient because the basis functions $\cu{\cos(m\varphi),\sin(m\varphi)}$ define pointed stars with sharp cusps for integers $\ab{m}>1$.  For odd $m$, these are $m$-pointed stars, with their parameterization tracing their shape twice (e.g., Fig.~\ref{fig:Examples} bottom), while for even $m$, these are $2m$-pointed stars, with their parameterization tracing them only once.

\subsection{Antipodal decomposition}

\begin{figure}
	\centering
	\includegraphics[width=\columnwidth]{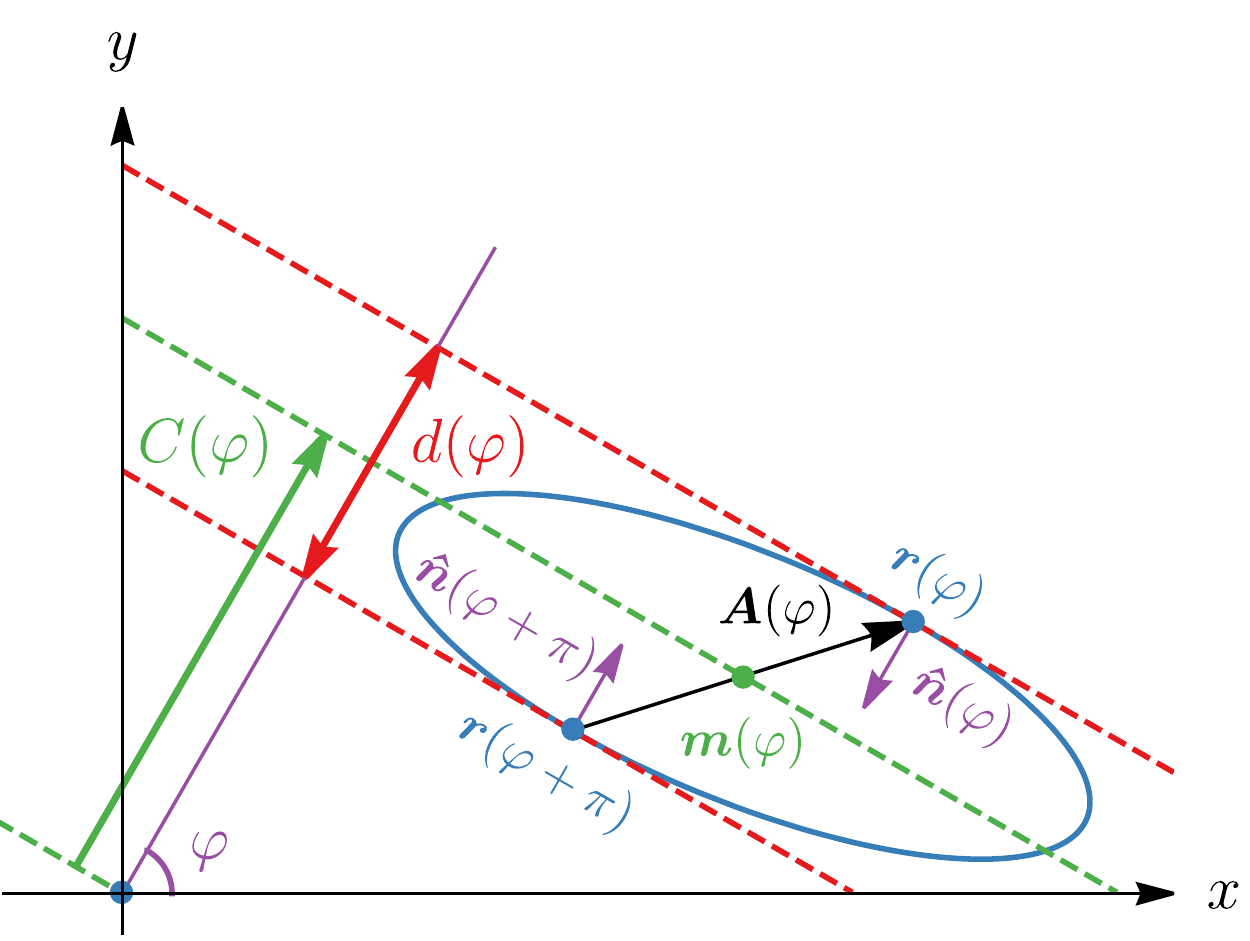}
	\caption{For a closed, convex curve, each angle $\varphi$ defines a pair of antipodal points $\bm{r}(\varphi)$ and $\bm{r}(\varphi+\pi)$ (blue dots).  In interferometry, one measures the projection of these points onto a line at angle $\varphi$ (purple).  We denote by $d(\varphi)$ the projected distance (red), which is the angle-dependent projected diameter, or width, of the shape.  The midpoint between the antipodes is denoted $\bm{m}(\varphi)$, and its projection $C(\varphi)$ is interpreted as an angle-dependent projected centroid (green).}
	\label{fig:Decomposition}
\end{figure}

A closed, convex curve admits a natural, well-defined notion of antipodal points: those whose tangent lines have the same slope.  In our framework, two antipodal points have normal angles $\varphi$ that differ by $\pi$.  For each normal angle $\varphi$, we introduce the antipodal displacement vector $\bm{A}$ and the midpoint position $\bm{m}$ \text{along the antipodal chord},
\begin{align}
	\bm{A}(\varphi)&=\bm{r}(\varphi)-\bm{r}(\varphi+\pi),\\
	\bm{m}(\varphi)&=\frac{\bm{r}(\varphi)+\bm{r}(\varphi+\pi)}{2}.
\end{align}

Each of these vector functions defines a closed curve on the plane.  The curve defined by $\bm{A}$ is symmetric under reflections through the origin (or equivalently, rotations by $\pi$), since $\bm{A}$ flips sign under $\varphi\to\varphi+\pi$.  It encodes the overall dimensions of the original curve and for this reason will be called its \textit{hull}.  The curve defined by $\bm{m}$ is in general displaced from the origin and always features sharp cusps, since $\bm{m}$ returns to itself after a lapse of only $\varphi$, during which time the continuous normal $\bm{\hat{n}}$ flips sign.  This \textit{midpoint curve} encodes the original curve's position on the plane as well as its fine features.  The parameterization $\bm{m}(\varphi)$ traces out the midpoint curve twice over the range $\varphi\in[0,2\pi)$, as illustrated in Fig.~\ref{fig:CuspyCurve}.  

\begin{figure}
	\centering
	\includegraphics[width=\columnwidth]{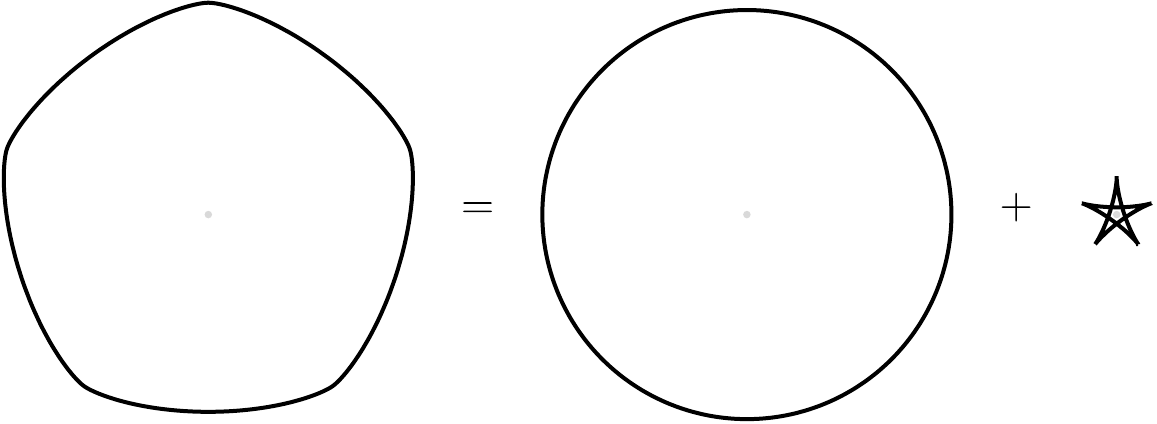}
	\caption{Illustration of the decomposition of a convex shape into its hull and midpoint curve.  The projected position function $f(\varphi)=1+\sin\pa{5\varphi}/28$ produces a rounded pentagon of constant width (left).  Its parity-even part $d(\varphi)=2$ defines a circle (middle), while its parity-odd part $C(\varphi)=\sin\pa{5\varphi}/28$ defines a sinestar (right).  We may therefore regard the rounded pentagon as the sum of a circle and a sinestar.}
	\label{fig:ShapeAddition}
\end{figure}

In other words, the notion of antipodal points naturally leads to a decomposition of any closed, convex curve into a symmetric, centered hull and a cuspy, displaced midpoint curve.  The full curve is recovered by vector addition of the hull and midpoint curves, identifying points that share a common normal angle.  Each point on the midpoint curve has two normal angles that differ by $\pi$, so that the full midpoint curve is thus included twice in the sum.  An example is illustrated in Fig.~\ref{fig:ShapeAddition}.

The projected positions of the hull and midpoint curves will be denoted $d$ and $C$, respectively,
\begin{align}
	d(\varphi)&=-\bm{A}\cdot\bm{\hat{n}}
	>0,\\
	C(\varphi)&=-\bm{m}\cdot\bm{\hat{n}}.
\end{align}
These are to be interpreted as the projected diameter and centroid, respectively (Fig.~\ref{fig:Decomposition}).  The projected diameter is always positive, while the projected centroid can take either sign.  The full projected position $f$ is decomposed into $d$ and $C$ as
\begin{align}
	\label{eq:AntipodalSplit}
	f(\varphi)=\frac{1}{2}d(\varphi)+C(\varphi),
\end{align}
or equivalently,
\begin{align}
	d(\varphi)&=f(\varphi)+f(\varphi+\pi),\\
	C(\varphi)&=\frac{1}{2}\br{f(\varphi)-f(\varphi+\pi))}.
\end{align}
That is, the projected diameter and centroid are the parity-even and parity-odd parts of the projected position, respectively.  This decomposition is very natural in the context of interferometry, where $d$ is encoded in the visibility amplitude, while $C$ appears as an overall phase (see Eq.~(3) in Ref.~\cite{Gralla2020}).

It is often convenient to regard $d(\varphi)$ and $C(\varphi)$ as being defined on the smaller range $\varphi\in[0,\pi)$.  That is, we may view a closed, convex curve as being specified by a single function $f(\varphi)$ ranging over the full circle $\varphi\in[0,2\pi)$, or equivalently, it can be repackaged into two functions $d(\varphi)$ and $C(\varphi)$ ranging over $\varphi\in[0,\pi)$.

Under shifts of the origin $\bm{r}\to\bm{r}+(X,Y)$, we have
\begin{align}
	d(\varphi)&\to d(\varphi),\\
	C(\varphi)&\to C(\varphi)+X\cos{\varphi}+Y\sin{\varphi}.
\end{align}
That is, all the coordinate origin information is contained in $C(\varphi)$, while $d(\varphi)$ remains invariant under translations, as required by its interpretation as the projected diameter.  To study shapes intrinsically, one might fix a canonical choice of $C$ by demanding that the dipole vanish, $\int C(\varphi)e^{\pm i\varphi}\ed\varphi=0$.  However, for interferometric measurements tracking absolute phase, the coordinate origin is ``known'' and the dipole of $C$ contains important information about where the curve is located on the image.

We can view a measurement of $d(\varphi)$ as a measurement of the hull of the underlying curve.  The remaining freedom is a choice of closed plane curve made of individually convex segments joined together at cusps.

\section{Examples}
\label{sec:Examples}

We now illustrate the framework of the previous section (Sec.~\ref{sec:ConvexCurves}) with a series of examples.

\subsection{The point}

The most trivial example is that of a curve that has degenerated to a single point $(X,Y)$.  Its projected position function is the pure dipole
\begin{align}
	\label{eq:Point}
	f_\mathrm{point}=X\cos{\varphi}+Y\sin{\varphi}.
\end{align}
This form may be added to any other $f(\varphi)$ to induce a translation.  It can also be used to add sharp corners to a shape by demanding that it hold exactly over some finite range of $\varphi$.  The parameterization then ``hovers'' at this point for that lapse of $\varphi$, during which time the normal vector still advances, creating a discontinuity in the normal (and tangent) vector of the shape.  The next example will illustrate this behavior.

\subsection{Curves of constant width}

Curves with $d'(\varphi)=0$ are generally called curves of constant width.  These are the curves whose hull is a circle.  The simplest example is the circle itself,
\begin{align}
	\label{eq:Circle}
	f_\mathrm{circle}=R,
\end{align}
where $R$ is the radius.  In this framework, one can easily construct other curves of constant width by adding an arbitrary function $C(\varphi)$ that is parity-odd,
\begin{align}
	f_\mathrm{c.w.}=R+C(\varphi).
\end{align}
For example, choosing $C(\varphi)=A\sin{3\varphi}$ with sufficiently small $A$ creates a triangular object with rounded sides and rounded corners.  Similarly, $C(\varphi)=A\sin{4\varphi}$ creates a rounded square, and so on.  The rounded pentagon with $C(\varphi)=A\sin{5\varphi}$ is shown in Fig.~\ref{fig:ShapeAddition}.

The most famous example of a non-circular constant-width curve is perhaps the Reuleaux triangle, which is composed of three circular arcs joined at $120^\circ$ angles.  This is a slightly subtle example in our formalism, as it contains sharp corners.  These are handled by using the point form \eqref{eq:Point} over the lapses of $\varphi$ corresponding to the jumps in $\bm{\hat{n}}$ at the corners.  The remainder of the shape consists of three circular arcs, each with curvature center at the opposite vertex of the triangle.  These are constructed using appropriate segments of the shifted circles $f=D+X\cos{\varphi}+Y\sin{\varphi}$ with appropriately chosen centers $(X,Y)$.  Choosing the geometric center of the Reuleaux triangle to be the origin and letting the top vertex lie on the positive $y$-axis, the projected position function is
\begin{align}
	\label{eq:Reuleaux}
	f_\mathrm{Rx}=D
	\begin{cases}
		1-\frac{1}{2\sqrt{3}}\sin{\varphi}-\frac{1}{2}\cos{\varphi}
		&0\le\varphi\le\frac{\pi}{3},\\
		\frac{1}{\sqrt{3}}\sin{\varphi}
		&\frac{\pi}{3}\le\varphi\le\frac{2\pi}{3},\\
		1-\frac{1}{2\sqrt{3}}\sin{\varphi}+\frac{1}{2}\cos{\varphi}
		&\frac{2\pi}{3}\le\varphi\le\pi,\\
		-\frac{1}{2\sqrt{3}}\sin{\varphi}-\frac{1}{2}\cos{\varphi}
		&\pi\le\varphi\le\frac{4\pi}{3},\\
		1+\frac{\sin\varphi}{\sqrt{3}}
		&\frac{4\pi}{3}\le\varphi\le\frac{5\pi}{3},\\
		\frac{1}{2}\cos{\varphi}-\frac{1}{2\sqrt{3}}\sin{\varphi}
		&\frac{5\pi}{3}\le\varphi\le2\pi. 
	\end{cases}
\end{align}

\begin{figure*}
	\centering
	\vspace{-25pt}
	\includegraphics[width=\textwidth]{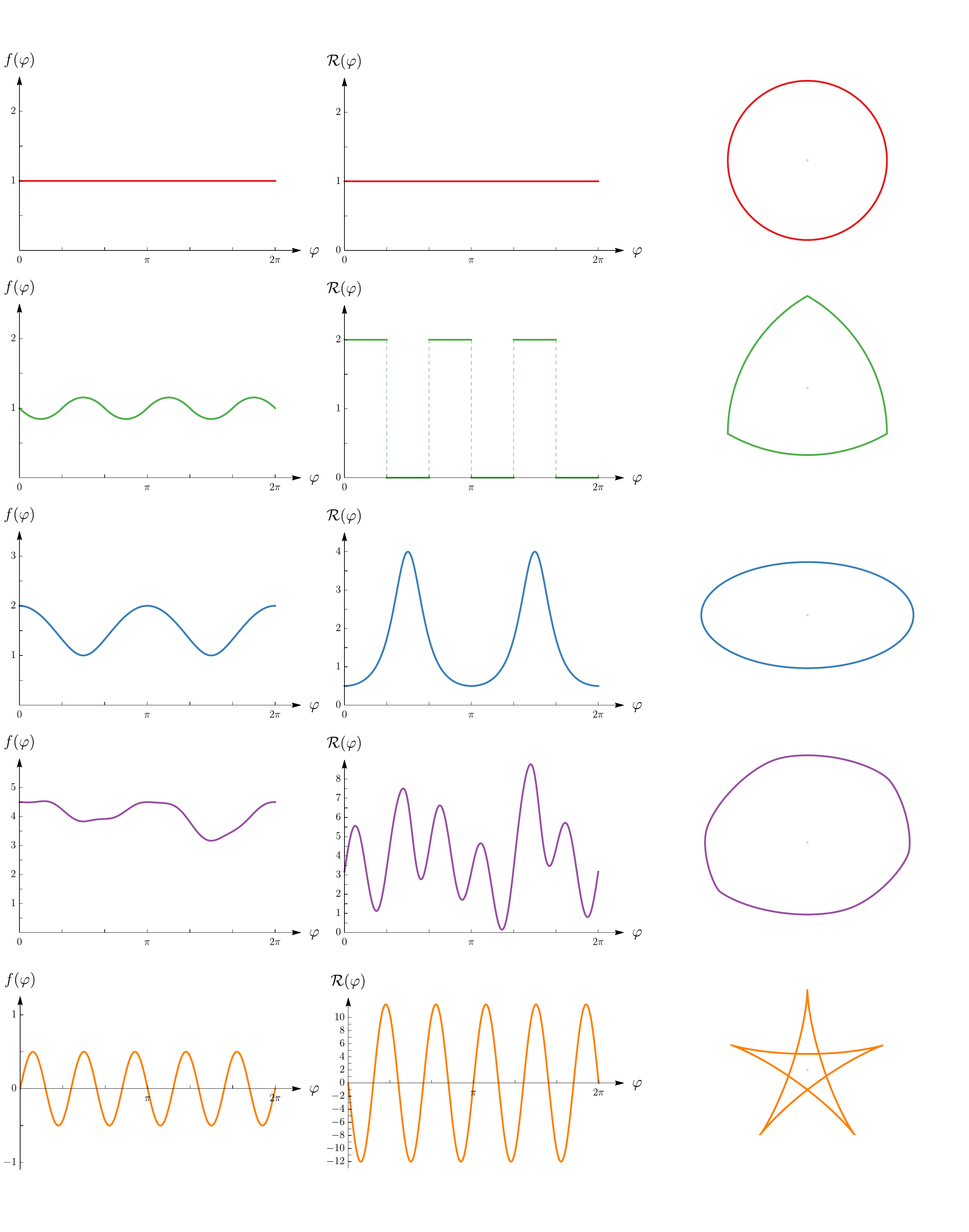}
	\vspace{-40pt}
	\caption{Illustration of the relationship between the projected position function $f(\varphi)$ (left) and its associated curve (right).  We also show the function $\mathcal{R}(\varphi)$ (middle), whose absolute value is the radius of curvature.  When this function passes through zero, the curve develops a cusp.  (The star shown in the bottom-right is traced twice as the parameter ranges over the full circle $\varphi\in[0,2\pi)$ and hence has only five cusps, despite the ten zero-crossings of $\mathcal{R}$.)  From top to bottom, we show the circle [Eq.~\eqref{eq:Circle} with $R=1$], the Reuleaux triangle [Eq.~\eqref{eq:Reuleaux} with $D=2$], the ellipse [Eq.~\eqref{eq:Ellipse} with $R_1=2$ and $R_2=1$], the pebble [Eq.~\eqref{eq:Pebble}], and the five-pointed sinestar $f(\varphi)=\sin{5\varphi}$.}
	\label{fig:Examples}
\end{figure*}

Although complex to write down, this function has a simple behavior, oscillating continuously (but not smoothly) between maxima and minima in a manner reminiscent of a pure sinusoid (Fig.~\ref{fig:Examples}).  It can be very closely approximated by a function of the form $A+B\sin{3\varphi}$, but it is not exactly of this form.

\subsection{Ellipse}

The parameterization of the ellipse by polar angle $\theta$ is
\begin{align}
	\label{eq:PolarEllipse}
	\bm{R}(\theta)=\pa{R_1\cos{\theta},R_2\sin{\theta}},
\end{align}
where $R_1$ and $R_2$ are the semi-axis lengths in the $x$ and $y$ directions.  The normal angle $\varphi(\theta)$ is found from Eq.~\eqref{eq:ParameterizedAngle}, which becomes
\begin{align}
	\tan{\varphi}=\frac{R_1}{R_2}\tan{\theta}. 
\end{align}
This relation is simple enough that it can be inverted as
\begin{align}
	\cos{\theta}&=\frac{R_1\cos{\varphi}}{\sqrt{R_1^2\cos^2{\varphi}+R_2^2\sin^2{\varphi}}},\\
	\sin{\theta}&=\frac{R_2\sin{\varphi}}{\sqrt{R_1^2\cos^2{\varphi}+R_2^2\sin^2{\varphi}}}.
\end{align}
Plugging into Eq.~\eqref{eq:PolarEllipse} with $\bm{r}(\varphi)=\bm{R}(\theta(\varphi))$ then yields the parameterization of the ellipse by normal angle,
\begin{align}
	x(\varphi)&=\frac{R_1^2\cos{\varphi}}{\sqrt{R_1^2\cos^2{\varphi}+R_2^2\sin^2{\varphi}}},\\
	y(\varphi)&=\frac{R_2^2\sin{\varphi}}{\sqrt{R_1^2\cos^2{\varphi}+R_2^2\sin^2{\varphi}}}.
\end{align}
The projected position is calculated from Eq.~\eqref{eq:ExplicitPosition} as
\begin{align}
	\label{eq:Ellipse}
	f_\mathrm{ellipse}=\sqrt{R_1^2\cos^2{\varphi}+R_2^2\sin^2{\varphi}},
\end{align}
which is invariant under antipodal exchange $\varphi\to\varphi+\pi$.  Hence, we find
\begin{align}
	d_\mathrm{ellipse}=2\sqrt{R_1^2\cos^2{\varphi}+R_2^2\sin^2{\varphi}},\quad
	C_\mathrm{ellipse}=0.
\end{align}
That is, the ellipse has antipodal symmetry.

\subsection{Circlipse}

Adding the projected position functions of two shapes produces a third shape corresponding to the vector sum of points that share a common normal angle.  A simple nontrivial example is the sum of a circle and an ellipse, or a ``circlipse'':
\begin{align}
	\label{eq:Circlipse}
	f_\mathrm{circlipse}=R_0+\sqrt{R_1^2\cos^2{\varphi}+R_2^2\sin^2{\varphi}}.
\end{align}
This is an antipodally symmetric shape with an oval form.  The semi-axes in the $x$ and $y$ directions are $R_0+|R_1|$ and $R_0+|R_2|$, respectively.  As discussed in Sec.~\ref{sec:CriticalCurve} below, the hull of the Kerr critical curve is closely approximated by a circlipse over the entire parameter space (see also Fig.~\ref{fig:CiriticalCusps}).  We are unaware of precious discussion of this shape, which arose in our studies of the Kerr critical curve.  Using the method of implicitization reviewed in App.~\ref{app:Implicitization}, we find that the circlipse is part of the vanishing locus of an order 8 polynomial, which does not appear to correspond to previously studied octic curves.

When $R_0=0$, the circlipse reduces to an ellipse, and when $R_1=R_2=0$, it becomes a circle.  If only one of $R_1$ or $R_2$ vanishes, then we obtain a singular limit that produces a ``racetrack'' shape,
\begin{align}
	\label{eq:Racetrack}
	f_\mathrm{race track}=R_0+R_2\ab{\cos{\varphi}}.
\end{align}
This function produces two half-circles of radius $R_0$ separated by a distance of $2R_2$, constructed by cutting a circle vertically through its center and separating the pieces horizontally.  Joining the two with straight lines produces a racetrack shape.  The circlipse \eqref{eq:Circlipse} is closed at any non-zero $R_1$, meaning that its $R_1\to0$ limit is indeed the full closed racetrack.  We may also take Eq.~\eqref{eq:Racetrack} to represent the full closed curve if we adopt the natural convention that kinks in $f(\varphi)$ are to represent straight lines.

\subsection{A cuspy triangle}

Whereas studying $d(\varphi)$ for the Kerr critical curve led us to consider the circlipse, studying $C(\varphi)$ led us to
\begin{align}
	\label{eq:Goldilocks}
	f_\mathrm{cuspy\,triangle}=\arcsin\pa{\chi\cos{\varphi}},
\end{align}
where $\chi\in\br{-1,1}$.  This shape is a cuspy triangle, ranging from small and equilateral as $\chi\to0$ to finite and isosceles as $|\chi|\to1$.  The limiting case $|\chi|=1$ contains a straight line encoded by kinks in $f$ at $\varphi\mod\pi=0$.  This parameterization also moves the triangle rightward with increasing $\chi$.  This cuspy triangle closely approximates the midpoint curve of the Kerr critical curve (Fig.~\ref{fig:CiriticalCusps} bottom), with the parameter $\chi$ an increasing function of dimensionless spin $a/M$. 

\subsection{Pebble}

As a final example we consider the function
\begin{align}
	\label{eq:Pebble}
	f_\mathrm{pebble}=3&+\sqrt{\pa{3/2}^2\cos^2{\varphi}+\pa{1/2}^2\sin^2{\varphi}}\nonumber\\
	&+\frac{1}{4}\sin^3{2\varphi}+\frac{1}{3}\sin^3{\varphi},
\end{align}
which produces an irregularly shaped ``pebble'' (Fig.~\ref{fig:Examples}).

\section{The Kerr critical curve}
\label{sec:CriticalCurve}

The Kerr critical curve \cite{Bardeen1973} is a theoretical closed curve on the image plane of a camera aimed at a black hole.  It is defined by the asymptotic arrival positions of photons that orbit the black hole arbitrarily many times before escaping to the camera.  The shape depends on the black hole spin parameter $a$ and the observer inclination $\theta_o$ relative to the spin axis, with the overall size set by the black hole mass $M$.  Successive images of bulk matter emission asymptote to the critical curve (e.g., Ref.~\cite{GrallaLupsasca2020a}).  In this section, we study the critical curve shape as encoded by its interferometric observables.  We discuss general features of the curve, derive analytic results in various limits, and give analytic formulae that faithfully approximate the shape over the entire parameter space.

\subsection{General Properties}

The original work of Bardeen \cite{Bardeen1973} provides a parametric formula for the critical curve.  In the notation of Refs.~\cite{GrallaLupsasca2020a,GrallaLupsasca2020b}, Bardeen's formula is
\begin{subequations}
\label{eq:BardeenCoordinates}
\begin{align}
	\label{eq:Alpha}
	\alpha(\tilde{r})&=-\frac{\lambda(\tilde{r})}{\sin{\theta_o}},\\
	\label{eq:Beta}
	\beta_\pm(\tilde{r})&=\pm\sqrt{\eta(\tilde{r})+a^2\cos^2{\theta_o}-\lambda(\tilde{r})^2\cot^2{\theta_o}},
\end{align}
\end{subequations}
with
\begin{subequations}
\label{eq:CriticalParameters}
\begin{align}
	\lambda(\tilde{r})&=a+\frac{\tilde{r}}{a}\br{\tilde{r}-\frac{2\pa{\tilde{r}^2-2M\tilde{r}+a^2}}{\tilde{r}-M}},\\
	\eta(\tilde{r})&=\frac{\tilde{r}^{3}}{a^2}\br{\frac{4M\pa{\tilde{r}^2-2M\tilde{r}+a^2}}{\pa{\tilde{r}-M}^2}-\tilde{r}}.
\end{align}
\end{subequations}
Here, $\alpha$ and $\beta$ are Cartesian ``screen coordinates'' (with units of $M$) describing the image.  We set $G=c=1$ and use the range $0<a<M$, treating the edges as limits.  The curve is parameterized by $\tilde{r}$ in two separate segments $\pm$, corresponding to its upper-half ($+$) and lower-half ($-$).  The orientation is clockwise in the upper-half plane and counter-clockwise in the lower-half plane (see, e.g., Fig.~3 in Ref.~\cite{GrallaLupsasca2020a}).

\begin{figure}
	\centering
	\includegraphics[width=\columnwidth]{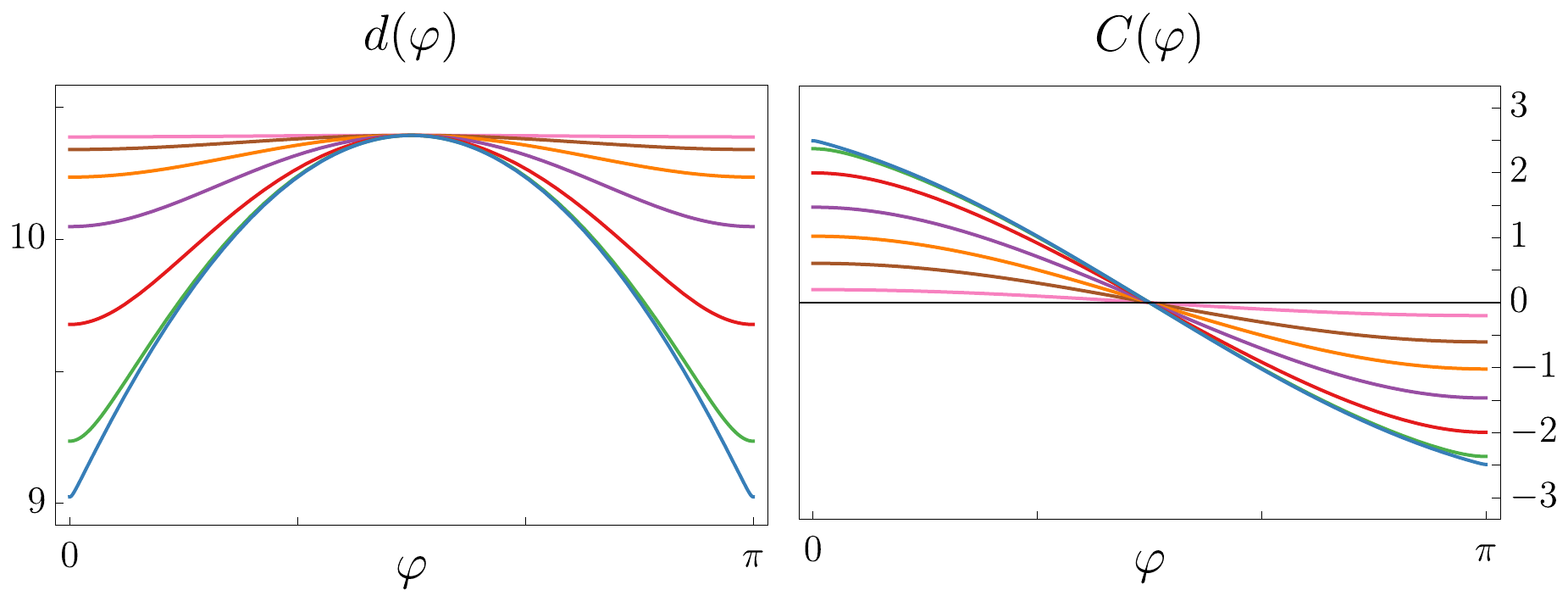}
	\includegraphics[width=\columnwidth]{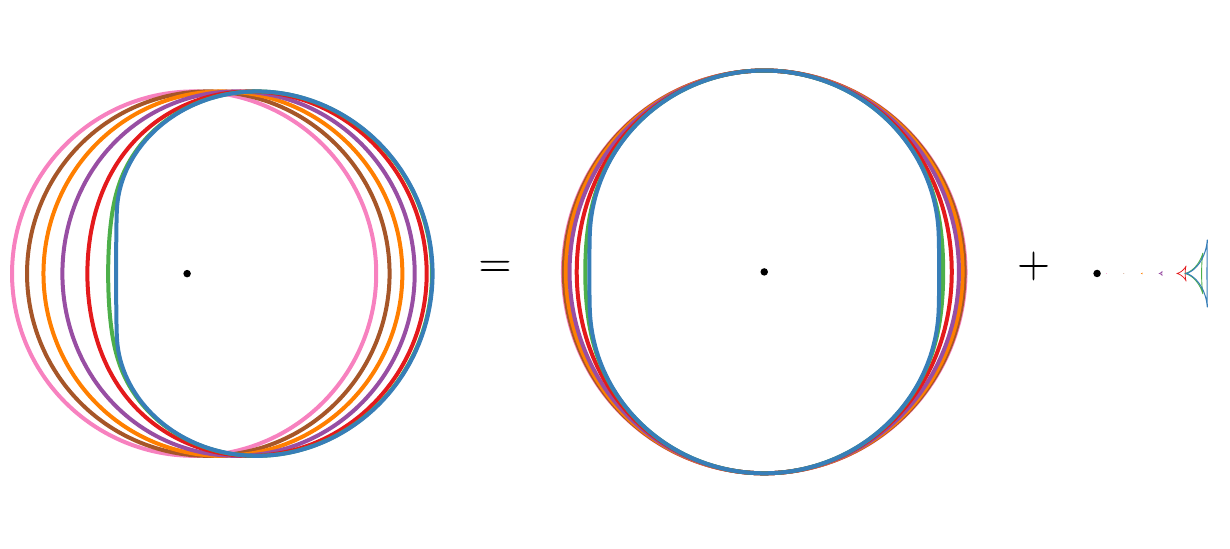}
	\includegraphics[width=\columnwidth]{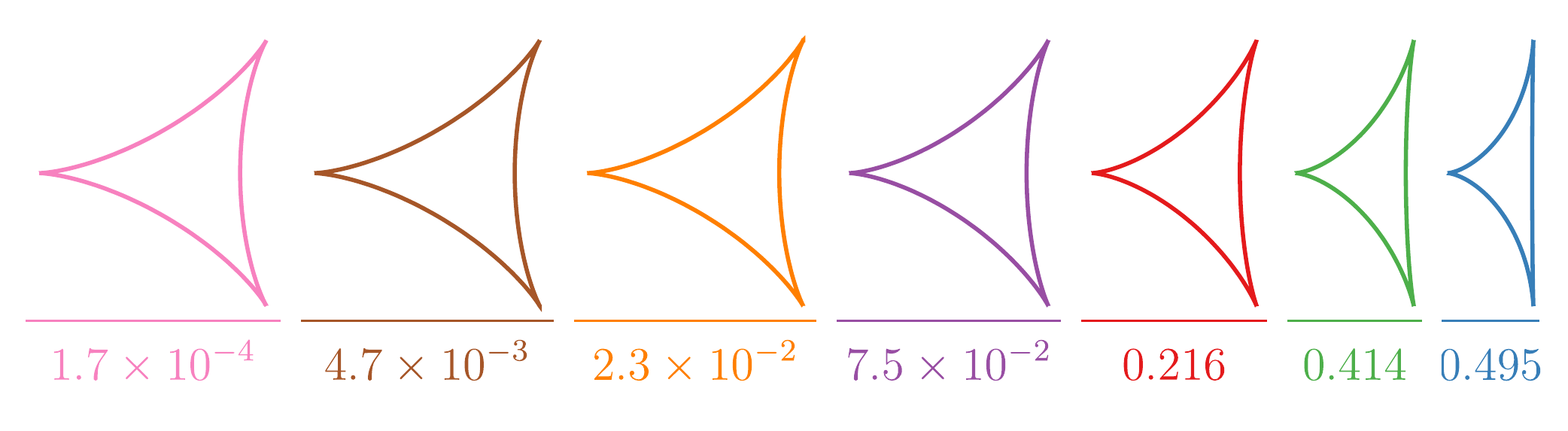}
	\caption{The shape decomposition of the critical curve.   The hull is closely approximated by a circlipse [Eq.~\eqref{eq:Circlipse}], while the midpoint curve is closely approximated by the cuspy triangle \eqref{eq:Goldilocks}.  The bottom panel shows a blow-up of these triangles with scale indicated in units of $M$.  We have set $M=1$ and chosen an equatorial observer $\theta_o=\pi/2$ with black hole spins \textcolor{pink}{10\%}, \textcolor{brown}{30\%}, \textcolor{orange}{50\%}, \textcolor{purple}{70\%}, \textcolor{red}{90\%}, \textcolor{green}{99\%}, \textcolor{blue}{99.99\%}.}
	\label{fig:CiriticalCusps}
\end{figure}

The parameter $\tilde{r}$ represents the radius at which photons orbit before reaching the detector.  The allowed range is the subset of the region $\tilde{r}\in\br{M,4M}$ for which $\beta$ remains real.  (Allowing $\tilde{r}$ to take any values defines a larger object, the \textit{critical locus}, that we study in App.~\ref{app:AlgebraicGeometry}.)  This range shrinks to zero at vanishing spin and/or inclination, where all photons orbit at the same radius.  Thus, this parameterization can only be used at non-zero spin and inclination.

The normal angle $\varphi$ is determined from Eq.~\eqref{eq:ParameterizedAngle} using Eqs.~\eqref{eq:BardeenCoordinates}.  After some simplification, we find that
\begin{align}
	\label{eq:CriticalTangent}
	\tan{\varphi(\tilde{r})}=\frac{\beta(\tilde{r})}{\alpha(\tilde{r})-a\sin{\theta_o}\pa{\frac{\tilde{r}+M}{\tilde{r}-M}}}.
\end{align}
Inverting for $\tilde{r}(\varphi)$ would provide $f(\varphi)$ analytically, but unfortunately, this requires solving a sextic polynomial.

Nonetheless, it is straightforward to use $\varphi(\tilde{r})$ to make parametric plots of $f(\varphi)$.  Defining $\arctan(x,y)\in(-\pi,\pi]$ to be the usual principal argument of the complex number $x+iy$, we have
\begin{align}
	\label{eq:CriticalNormalAngle}
	\varphi_\pm(\tilde{r})=\arctan\pa{\alpha(\tilde{r})-a\sin{\theta_o}\pa{\frac{\tilde{r}+M}{\tilde{r}-M}},\beta_\pm(\tilde{r})},
\end{align}
where $\varphi_\pm$ is the normal angle in the upper-half ($+$) and lower-half ($-$) planes.  This formula holds modulo $2\pi$; to obtain our canonical range $\varphi\in[0,2\pi)$, one must add $2\pi$ to $\varphi_-$.

Together, Eqs.~\eqref{eq:ExplicitPosition}, \eqref{eq:BardeenCoordinates}, and \eqref{eq:CriticalNormalAngle} imply that $f(\varphi)$ can be plotted as the two-segment parametric curve $(\varphi_\pm(\tilde{r}),f(\varphi_\pm(\tilde{r}))$.  However, particular values of $f(\varphi)$ must still be extracted numerically (say, by a graphical method using the plot data).  Similarly, the decomposition into $d(\varphi)$ and $C(\varphi)$ requires a numerical method.

In Fig.~\ref{fig:CiriticalCusps}, we illustrate the properties of the Kerr critical curve in the language of this paper.  We pick the equatorial observer $\theta_o=\pi/2$ for which the shape variation with spin is greatest.  Similar patterns are seen at smaller inclination angles, albeit with less overall variation.

In studying the projected position function of the critical curve, we have found some simple functional forms that approximate it to remarkably high accuracy.  In particular, the even part $d(\varphi)$ is well-described by the circlipse form shown above in Eq.~\eqref{eq:Circlipse}, while the odd part $C(\varphi)$ is well-described by the cuspy triangle \eqref{eq:Goldilocks} together with a horizontal translation \eqref{eq:Point}.  The precise critical curve is therefore well-described by the projected position function
\begin{align}
	f(\varphi)=R_0&+\sqrt{R_1^2\sin^2{\varphi}+R_2^2\cos^2{\varphi}}\nonumber\\
	\label{eq:Goldylipse}
	&+\pa{X-\chi}\cos{\varphi}+\arcsin\pa{\chi\cos{\varphi}}.
\end{align}
We will call this a \textit{phoval}, for ``photon ring oval''. 

We have chosen the translation parameter $X$ such that $X\cos{\varphi}$ is the dipole term in the Fourier cosine series of this function.  The intrinsic shape of the phoval has four parameters, namely: three nonnegative radii $R_0$, $R_1$, $R_2$, together with an asymmetry parameter $-1\le\chi\le1$.  Provided the parameters are chosen such that the shape is convex, the phoval has two preferred axes that intersect it orthogonally.  In the form \eqref{eq:Goldylipse} with $X=0$, these are the $x$ and $y$ axes.  The horizontal radius is $R_0+R_1$ and the vertical radius is $R_0+R_2$.  The relative size of $R_1$ and $R_2$ determines how flattened or rounded the associated edges are, and the parameter $\chi$ introduces a horizontal asymmetry in this roundedness.

The phoval provides an excellent fit to the Kerr critical curve.  For each choice of black hole spin and observer inclination, we plot the critical curve projected position function $f(\varphi)$ as described above.  We then fit the functional form \eqref{eq:Goldylipse} to the plot data, finding best-fit parameters and normalized root-mean-square (RMS) residuals, defined as the average of the squared-deviation divided by the span, i.e., $\langle(\delta f)^2\rangle/(f_{\rm max}-f_{\rm min})$.  Repeating this procedure over the whole parameter space, sampled uniformly in spin $a\in(0,M)$ and inclination $\theta_o\in(0,\pi/2]$ (in practice, we start at values very close to the edges of these ranges), we find a median normalized RMS deviation of $10^{-5}$.  The largest residuals are a few times $10^{-3}$, occurring in the extremal limit $a\to M$.  (We have computed for $a$ as large as $a=.9999M$.)  That is, the fit works to a part in $10^5$ over the vast majority of the parameter space, and to a part in $10^3$ near extremality.

Very recently, Ref.~\cite{Farah2020} showed that the critical curve may also be approximated to great accuracy by a lima\c{c}on curve, building on an older observation of Ref.~\cite{deVries2003}.  A direct comparison of the fit quality is difficult, since the fit diagnostics differ.  In particular, our diagnostic is tied to the interferometric signature $f(\varphi)$, whereas theirs is tied to the radial distance from an origin on the image plane.

\begin{figure}
	\centering
	\includegraphics[width=\columnwidth]{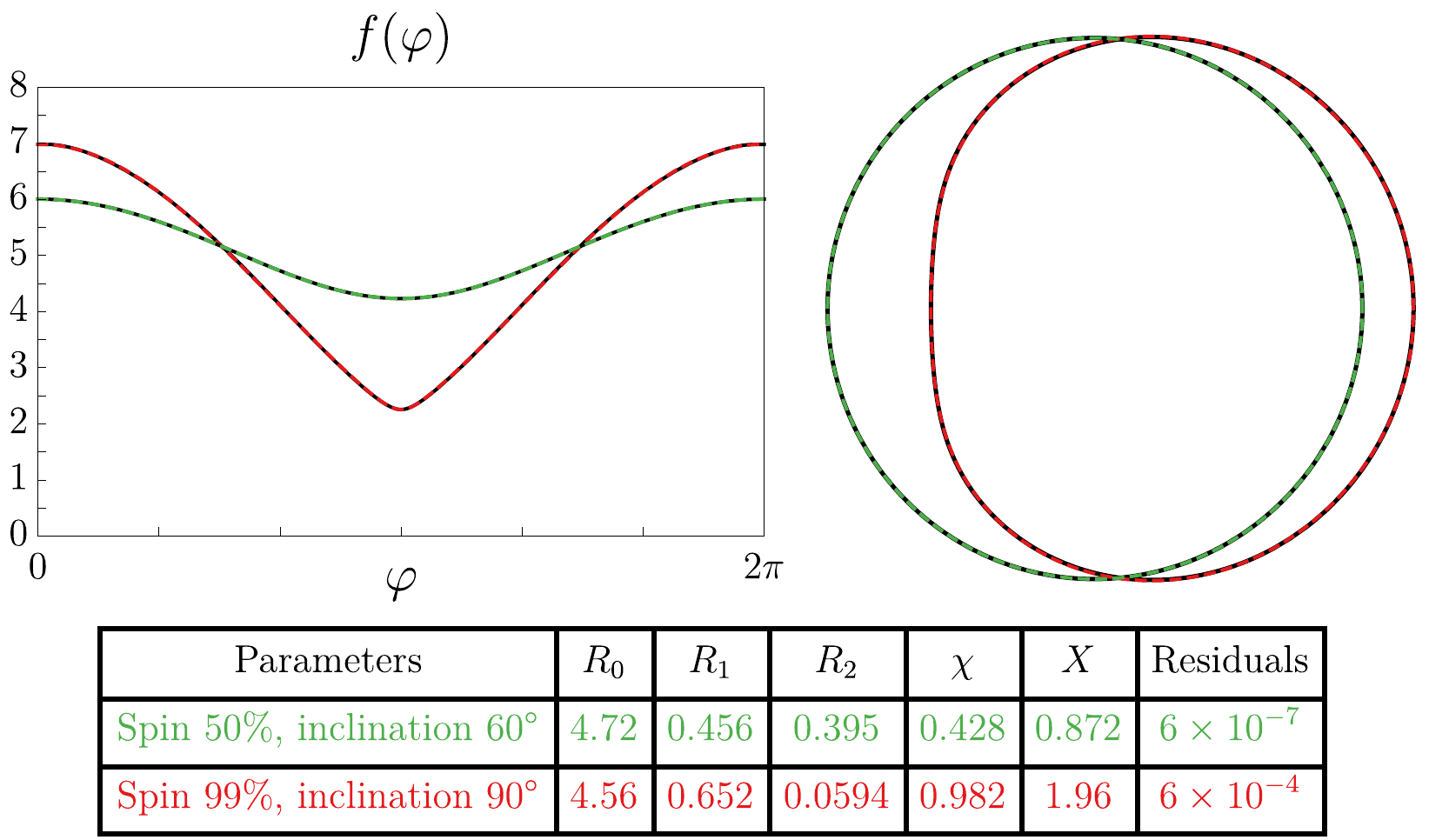}
	\caption{The Kerr critical curve as a phoval.  We show two examples of fitting the phoval shape (dashed curves) to critical curve (solid).  The resulting phoval is visually indistinguishable from the critical curve, both in terms of the projected position function (left) and the image plane curve (right).  The parameters and normalized RMS residuals are displayed in the table.  We have set $M=1$ in this figure.}
	\label{fig:Goldylipse}
\end{figure}

\subsection{Expansion in small spin or inclination}

At zero spin or inclination, the critical curve is precisely circular, as required by symmetry.  It remains circular up to second order in spin, when it takes the shape of an ellipse \cite{Bozza2006}.  Here, we reproduce this small-spin result, correcting some errors in the formulas for the ellipse parameters.  We also provide the analogous result at small inclination, proving that the critical curve is also an ellipse at second order in inclination (for any spin), and finding closed-form expressions for its parameters.

As discussed below Eqs.~\eqref{eq:CriticalParameters}, the parameter $\tilde{r}$ breaks down in the limit of small spin or small inclination.  Studying these limits requires an alternative expression of the critical curve.  We can eliminate the problematic parameter by using the formula
\begin{align}
	\label{eq:Radius}
	\tilde{r}&=M+2M\triangle\cos\br{\frac{1}{3}\arccos\pa{\frac{1-\frac{a^2}{M^2}}{\triangle^3}}},\nonumber\\
	&\triangle=\sqrt{1-\frac{a(a-\alpha\sin{\theta_o})}{3M^2}},
\end{align}
which follows from the $k=0$ case\footnote{The other inversions $k=1$ and $k=2$, reproduced in Eq.~\eqref{eq:CubicRoots} below, are not needed to recover the critical curve at sufficiently small spin or inclination.  See Fig.~1 of Ref.~\cite{GrallaLupsasca2020a} and Eqs.~(123)--(125) of the revised arXiv version of Ref.~\cite{GrallaLupsasca2020b}.} of Eq.~(122) in Ref.~\cite{GrallaLupsasca2020b} [wherein $\tilde{\lambda}=\lambda(\tilde{r})$], along with Eq.~\eqref{eq:Alpha} above.  This expresses $\tilde{r}$ in terms of $\alpha$, which is one of the screen coordinates.  Since the curve itself remains finite in the limit, the range of $\alpha$ must similarly remain finite, indicating that it can be used as a parameter in these limits.

That is, we may now discuss the critical curve as a pair of ordinary functions $\beta_\pm(\alpha)$.  Combining Eqs.~\eqref{eq:Alpha} and \eqref{eq:Beta} gives
\begin{align}
	\beta_\pm(\alpha)=\pm\sqrt{\eta(\tilde{r}(\alpha))+\pa{a^2-\alpha^2}\cos^2{\theta_o}},
\end{align}
with $\tilde{r}(\alpha)$ provided by Eq.~\eqref{eq:Radius}.  Another way of organizing this information is to write
\begin{align}
	\label{eq:CriticalCurve}
	\alpha^2+\beta^2=a^2\cos^2{\theta_o}+F(a,\alpha\sin{\theta_o}),
\end{align}
where
\begin{align}
	\label{eq:F}
	F=\frac{\tilde{r}^{3}}{a^2}\br{\frac{4M\pa{\tilde{r}^2-2M\tilde{r}+a^2}}{\pa{\tilde{r}-M}^2}-\tilde{r}}+\alpha^2\sin^2{\theta_o},
\end{align}
with $\tilde{r}$ given by Eq.~\eqref{eq:Radius}.  In regarding $F$ as a function of $a$ and $\alpha\sin{\theta_o}$, we are suppressing dependence on the overall scale $M$.

This function $F$ has a regular expansion at small spin (for any inclination) as well as at small inclination (for any spin).  We may imagine expanding to some fixed order in either parameter and determining the critical curve from Eq.~\eqref{eq:CriticalCurve}.  The functional form of \eqref{eq:CriticalCurve} shows immediately that, at quadratic order in either small parameter, the critical curve is determined by the roots of a quadratic polynomial in $\alpha$ and $\beta$.  This is on general grounds a conic section, and since the critical curve is closed, it must be an ellipse.  That is, Eq.~\eqref{eq:CriticalCurve} immediately shows that the critical curve is an ellipse at small spin and/or inclination.

It is rather straightforward to extract the parameters of the ellipse by performing the expansion in $F$.  When expanding in spin, we find 
\begin{align}
	F=27M^2+\pa{4\alpha\sin{\theta_o}}a-\br{\frac{\alpha^2\sin^2{\theta_o}}{9M^2}+4}a^2+\O{a^3}.
\end{align}
Let us first consider the shape at linear order.  Plugging into Eq.~\eqref{eq:CriticalCurve}, we obtain
\begin{align}
	\pa{\alpha-2a\sin{\theta_o}}^2+\beta^2=27M^2+\O{a^2},
\end{align}
demonstrating that the curve is a circle of radius $3\sqrt{3}M$ centered at $\alpha=2a\sin{\theta_o}$.  At quadratic order, we find
\begin{align}
	\pa{\frac{\alpha-2a\sin{\theta_o}}{1-\frac{a^2\sin^2{\theta_o}}{18M^2}}}^2+\beta^2=27M^2-3a^2\cos^2{\theta_o}+\O{a^3}.
\end{align}
This can be put in the canonical form of an ellipse,
\begin{align}
	\label{eq:CanonicalEllipse}
	\pa{\frac{\alpha-\alpha_0}{R_1}}^2+\pa{\frac{\beta}{R_2}}^2=1,
\end{align}
with
\begin{align}
	\label{eq:CenterOffset}
	\alpha_0&=2a\sin{\theta_o}+\O{a^3},\\
	\label{eq:RadiusCorrection1}
	R_1&=3\sqrt{3}M\pa{1-\frac{a^2}{18M^2}}+\O{a^3},\\
	\label{eq:RadiusCorrection2}
	R_2&=3\sqrt{3}M\pa{1-\frac{a^2\cos^2{\theta_o}}{18M^2}}+\O{a^3}.
\end{align}
The shape of the critical curve of a slowly spinning black hole was previously analyzed in Ref.~\cite{Bozza2006}.  Our Eq.~\eqref{eq:CenterOffset} is in agreement with their Eq.~(22), but our Eqs.~\eqref{eq:RadiusCorrection1} and \eqref{eq:RadiusCorrection2} differ from their Eqs.~(23) and (24) by some simple factors.  We have checked numerically that our results are correct.

The ellipse parameters at small inclination (for any spin) can be found by the same method: expand $F$ to quadratic order in $\alpha \sin \theta_o$ and compare to the canonical form.  This produces rather unwieldy analytic formulas that we do not display here.  In App.~\ref{app:AlgebraicGeometry}, we find somewhat less unwieldy formulas using an algebraic geometry approach.  In particular, we are able to write the ellipse parameters as rational functions of the spin $a$ and the radius $\tilde{b}(a)$ of the zero-inclination critical curve [see Eqs.~\eqref{eq:ParametersStart}--\eqref{eq:ParametersEnd} below].

\subsection{Extremal limit}
\label{sec:ExtremalCriticalCurve}

\begin{figure*}
	\centering
	\includegraphics[width=\textwidth]{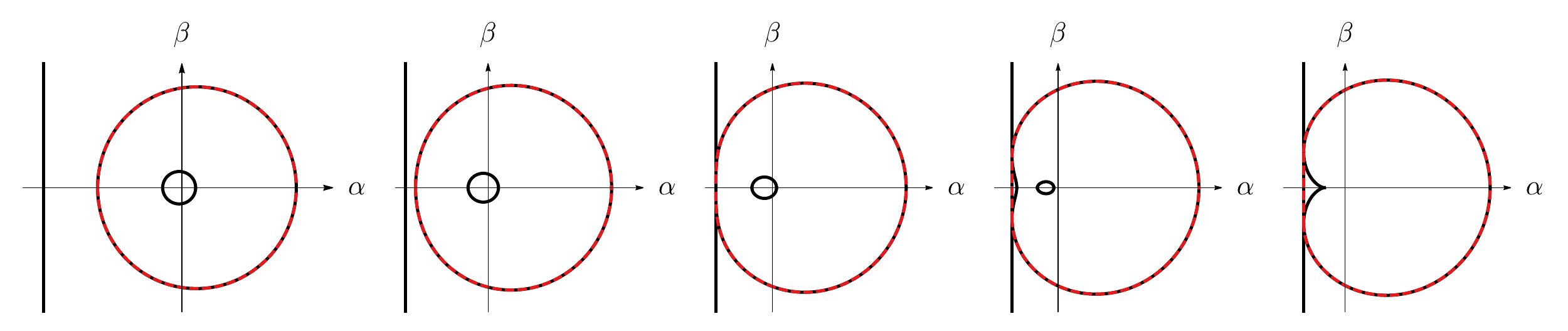}
	\caption{The extremal critical locus (black) consists of a vertical line together with a pair of Cartesian ovals.  The physical critical curve (dashed, red) is a subset of this locus that can be succinctly described as the convex hull of the ovals.  At small observer inclinations, the ovals are nearly circular and concentric about the origin, and the line is far away to the left.  As the inclination is increased ($\sin{\theta_o}=0.3,0.5,\sqrt{3}-1\approx0.732,0.9,1$ from left to right), the line moves towards the larger oval, which flattens to meet it; the smaller oval moves leftward.  The line touches the oval at the critical inclination $\sin{\theta_o}=\sqrt{3}-1$, where the NHEK spacetime becomes visible \cite{Gralla2018}.  After the kiss, the line remains attached to the oval, which develops a dimple as it smiles in response.  The inner oval continues to move left, eventually merging with the dimple at the cusp of a cardioid.}
	\label{fig:ExtremalCurves}
\end{figure*}

Finally, we consider the extremal ($a\to M$) limit of the critical curve.  In this case, the curve picks up a straight line (the ``NHEKline'') that can be attributed to the near-horizon geometry \cite{Bardeen1973,Gralla2018}.  This straight line means that our description in terms of functions $\beta_\pm(\alpha)$ breaks down as $a\to M$.  However, it is still instructive to analyze the portion of the curve for which it remains valid.  This will lead us to the result that the extremal critical curve is the convex hull of a pair of Cartesian ovals.

As noted in the previous section, the substitution of $\tilde{r}(\alpha)$ [Eq.~\eqref{eq:Radius}] into $\beta_\pm(\tilde{r})$ [Eq.~\eqref{eq:BardeenCoordinates}] provides a valid parameterization $\beta_\pm(\alpha)$ of the entire critical curve only for sufficiently small spin and inclination.  Above a certain threshold in these parameters, we recover only a part of the critical curve with this substitution (with the other portion requiring another inversion of $\lambda(\tilde{r})$ given in Eq.~\eqref{eq:CubicRoots} below).  We will see that in the extremal limit $a\to M$ (for any inclination), this portion gives precisely the non-straight part, i.e., the curve minus its NHEKline.  Thus, to study this portion, we set $a=M$ in Eqs.~\eqref{eq:Radius} and \eqref{eq:F}, finding the comparatively simple expressions
\begin{align}
	\tilde{r}=M\pa{1+\sqrt{2+\frac{\alpha\sin{\theta_o}}{M}}},
\end{align}
and
\begin{align}
	F=\alpha^2\sin^2{\theta_o}^2-\frac{\tilde{r}^3\pa{\tilde{r}-4M}}{M^2},
\end{align}
so that Eq.~\eqref{eq:CriticalCurve} becomes
\begin{align}
	\label{eq:ExtremalCriticalCurve}
	\pa{\alpha-M\sin{\theta_o}}^2+\beta^2-12M^2=8M^2\sqrt{2+\frac{\alpha\sin{\theta_o}}{M}}.
\end{align}
This curve fails to close for $\sin{\theta_o}>\sqrt{3}-1$.  Comparing with Eq.~(A6) of Ref.~\cite{Gralla2018}, we see that the missing segment is precisely the NHEKline
$\alpha=-2M\csc{\theta_o}$ with endpoints $\beta_\pm=\pm M\sqrt{3+\cos^2{\theta_o}-4\cot^2{\theta_o}}$, as claimed above.  To elucidate the shape of the extremal critical curve, we may square both sides of Eq.~\eqref{eq:ExtremalCriticalCurve} to obtain a quartic equation $\mathcal{E}=0$, with
\begin{align}
	\label{eq:CartesianOvals}
	\mathcal{E}=&\br{\pa{\alpha-M\sin{\theta_o}}^2+\beta^2-12M^2}^2\nonumber\\
	&\quad-64M^3\pa{2M+\alpha\sin{\theta_o}}.
\end{align}
After a simple translation in $\alpha$ to Cartesian coordinates $(x,y)=(\alpha-M\sin{\theta_o},\beta)$, we recognize $\mathcal{E}$ to be the defining equation for a pair of Cartesian ovals,
\begin{align}
	\pa{x^2+y^2}^2+k\pa{x^2+y^2}+lx+m=0,
\end{align}
with parameters $k=-24M^2$, $l=-64M^3\sin{\theta_o}$, and $m=16M^4\pa{1-4\sin^2{\theta_o}}$.\footnote{Under a different translation, it is also possible to put $\mathcal{E}$ in the form $\br{\pa{1-m^2}\pa{X^2+Y^2}+2m^2cX+A^2-m^2c^2}^2=4A^2\pa{X^2+Y^2}$ that is more traditionally given for the Cartesian oval, at the cost of complicated expressions for the parameters $m$, $c$, and $A$.}  Squaring Eq.~\eqref{eq:ExtremalCriticalCurve} introduced an unphysical curve in Eq.~\eqref{eq:CartesianOvals}: solving for $\beta$ gives
\begin{align}
	\label{eq:Solutions}
	\beta^2=12M^2-\pa{\alpha-M\sin{\theta}}^2\pm8M^{3/2}\sqrt{2M+\alpha\sin{\theta}},
\end{align}
with only the plus sign corresponding to the physical critical curve \eqref{eq:ExtremalCriticalCurve} (more precisely, its non-straight portion).

The full set of solutions \eqref{eq:Solutions} describes a pair of Cartesian ovals.  When $\sin{\theta_o}\le\sqrt{3}-1$, the ovals are both convex and the outer one is the full critical curve.  When $\sin{\theta_o}>\sqrt{3}-1$, the outer oval becomes non-convex, and the critical curve is formed by adjoining the NHEKline to the portion of the oval described by Eq.~\eqref{eq:ExtremalCriticalCurve}.  That is, at any observer inclination, the extremal critical curve is the convex hull of a Cartesian oval.

It is entertaining to ponder how a shape first studied by Descartes in 1637 is thereby embedded into the Kerr metric of general relativity.

In App.~\ref{app:AlgebraicGeometry}, we recover the extremal critical curve in a different way, by viewing it as a subset of a larger ``critical locus''.  The critical locus includes additional curves corresponding to unphysical values of the parameter $\tilde{r}$, but as a result can be described as an algebraic variety (in this case, the vanishing locus of a single polynomial).  The extremal critical locus is composed of the full pair of Cartesian ovals \eqref{eq:CartesianOvals} together with a straight line, as illustrated in Fig.~\ref{fig:ExtremalCurves}.\footnote{The critical locus at generic spin does not contain a Cartesian oval.  This follows from the fact demonstrated in App.~\ref{app:AlgebraicGeometry} that its defining polynomial $\mathcal{C}_o$ is of the form $\prod_{i=1}^3\br{\beta^2-f_i(\alpha)}$, with none of the $f_i$ a polynomial [see Eqs.~\eqref{eq:CriticalPolynomial}, \eqref{eq:CriticalPieces}, \eqref{eq:Factorization}].  At extremality, one of these factors becomes a quadratic describing a simple line (with multiplicity two), implying that the other two factors necessarily multiply to form a quartic polynomial---the Cartesian ovals \eqref{eq:CartesianOvals}.}

\subsection{Extremal, equatorial shape}

Recently, Ref.~\cite{Farah2020} noted that the extremal critical curve of an equatorial observer is the convex hull of a cardioid.  This fact is apparent in the rightmost panel of Fig.~\ref{fig:ExtremalCurves}, given that the fully  degenerate case of a Cartesian oval is known to be a cardioid.  We may also see it algebraically by noting that for $\theta_o=\pi/2$, the polynomial \eqref{eq:CartesianOvals} defining the Cartesian ovals can be recast in the canonical form of a cardioid,
\begin{align}
	\label{eq:Cardioid}
	\pa{X^2+Y^2}^2-4AX\pa{X^2+Y^2}=4A^2Y^2,  
\end{align}
with $X=\alpha+M$, $Y=\beta$, and $A=2M$.

The normal-angle parameterization for the convex hull of the cardioid (i.e., the extremal, equatorial critical curve) can be computed in closed form.  We begin with the parameterization of the full cardioid \eqref{eq:Cardioid} by polar angle $\theta\in[0,2\pi)$,
\begin{subequations}
\label{eq:PolarExtremalCurve}
\begin{align}
	\alpha(\theta)&=M\pa{1-4\cos{\theta}+2\cos{2\theta}},\\
	\beta(\theta)&=M\pa{-4\sin{\theta}+2\sin{2\theta}}.
\end{align}
\end{subequations}
The angle $\varphi(\theta)$ is found from Eq.~\eqref{eq:ParameterizedAngle}, which becomes
\begin{align}
	\tan{\varphi}=-\cot{\frac{3\theta}{2}}. 
\end{align}
This relation is simple enough that it can be inverted as
\begin{align}
	\theta_I(\varphi)=\frac{2\varphi+\pa{2I+1}\pi}{3},\quad
	I\in\mathbb{Z}.
\end{align}
Since the cardioid is not a closed, convex curve, we must resort to the general framework of Sec.~\ref{sec:GeneralCurves} to describe its shape.  Plugging into Eqs.~\eqref{eq:PolarExtremalCurve} with $\bm{r}(\varphi)=\bm{R}(\theta(\varphi))$ then yields the parameterization by $\varphi\in[0,\pi)$ of the edge-on extremal cardioid in three segments with $I\in\cu{-1,0,1}$,
\begin{align}
	\alpha_I(\varphi)=\alpha(\theta_I(\varphi)),\quad
	\beta_I(\varphi)=\beta(\theta_I(\varphi)).
\end{align}
Equivalently, these segments can be stitched into a single curve with $\varphi\in[-\pi,2\pi)$,
\begin{subequations}
\label{eq:ExtremeCurve}
\begin{align}
	\alpha(\varphi)&=M\pa{1+4\cos{\frac{2\varphi}{3}}+2\cos{\frac{4\varphi}{3}}},\\
	\beta(\varphi)&=M\pa{4\sin{\frac{2\varphi}{3}}+2\sin{\frac{4\varphi}{3}}}.
\end{align}
\end{subequations}
Discarding the $I=-1$ piece results in an open, convex curve consisting of two segments $I=1$ (``top'') and $I=0$ (``bottom''), whose convex hull is the critical curve.  This is equivalent to restricting the parameterization \eqref{eq:ExtremeCurve} to the circle $\varphi\in[-\pi,\pi)$.

The projected position on the range $\varphi\in[-\pi,\pi)$ is then calculated from Eq.~\eqref{eq:ExplicitPosition} as
\begin{align}
	f(\varphi)=M\pa{\cos{\varphi}+6\cos{\frac{\varphi}{3}}}.
\end{align}
By making this function $2\pi$-periodic, we can extend it to our canonical range $\varphi\in[0,2\pi)$, on which it becomes
\begin{align}
	f(\varphi)=M\cos{\varphi}+
	\begin{cases}
		6M\cos{\frac{\varphi}{3}},
		&0\le\varphi\le\pi,\\
		-6M\cos{\frac{\varphi+\pi}{3}}
		&\pi\le\varphi\le2\pi.
	\end{cases}
\end{align}
This formula gives the exact projected position function of the extremal, equatorial critical curve.  The kink at $\varphi=\pi$ represents the NHEKline.

\subsection{Projected position for equatorial critical curves}

We can in fact analytically compute the projected position of the critical curve for any equatorial observer, not just at extremal spin.  This is possible because the sextic polynomial discussed below Eq.~\eqref{eq:CriticalTangent} degenerates to a cubic when $\theta_o=\pi/2$.

More explicitly, when $\theta_o=\pi/2$, Eq.~\eqref{eq:CriticalTangent} reduces to
\begin{align}
	\cos^2{\varphi(\tilde{r})}=\frac{\tilde{r}\pa{\tilde{r}-3M}^2}{4Ma^2},
\end{align}
which is a cubic equation in $\tilde{r}$ with positive discriminant
\begin{align}
	\triangle_3=\frac{3^3\cos^2{\varphi}}{2^4a^6}\pa{1-\frac{a^2}{M^2}\cos^2{\varphi}}
	>0.
\end{align}
Hence, it has three real roots given for $k\in\cu{0,1,2}$ by
\begin{align}
	\tilde{r}_{(k)}(\varphi)=3M+\frac{a\cos{\varphi}}{\cos\br{\frac{1}{3}\arccos\pa{\frac{a}{M}\cos{\varphi}}-\frac{2\pi k}{3}}}.
\end{align}
Plugging the root $\tilde{r}_{(0)}(\varphi)$ into Eqs.~\eqref{eq:BardeenCoordinates} yields the parameterization $(\alpha(\varphi),\beta_\pm(\varphi))$ by normal angle $\varphi$, where one is to choose $+$ for $\varphi\in[0,\pi)$ and $-$ for $\varphi\in[\pi,2\pi)$.\footnote{On the range $\varphi\in[0,\pi)$, the root $\tilde{r}_{(2)}(\varphi)$ yields a parameterization with the opposite (clockwise) orientation, while the root $\tilde{r}_{(1)}(\varphi)$ traces twice over the upper half of the curve defined by mapping the $(\lambda,\eta)$-space curve $\mathcal{C}_-$ \cite{GrallaLupsasca2020b} onto the image plane via Eqs.~\eqref{eq:BardeenCoordinates}.}  The projected position is then calculated from Eq.~\eqref{eq:ExplicitPosition} as
\begin{align}
	f(\varphi)&=\alpha\pa{\tilde{r}_{(0)}(\varphi)}\cos{\varphi}+\beta\pa{\tilde{r}_{(0)}(\varphi)}\sin{\varphi},
\end{align}
with the choice of sign $\pm$ in $\beta(\tilde{r})$ given by $\mathrm{sign}\pa{\pi-\varphi}$.

\section*{Acknowledgements}

This work was supported in part by NSF grant PHY-1752809 to the University of Arizona.  AL acknowledges the Jacob Goldfield Foundation.

\appendix

\section{Perimeter}
\label{app:Perimeter}

The Cauchy surface area theorem states (in two dimensions) that the perimeter of a closed, convex curve is the average of its projected diameters.  The framework of Sec.~\ref{sec:ConvexCurves} provides an elegant proof of this fact.  Using Eq.~\eqref{eq:ConvexParameterization}, we have 
\begin{align}
	P&=\int_0^{2\pi}\sqrt{x'(\varphi)^2+y'(\varphi)^2}\ed\varphi\\
	&=\int_0^{2\pi}\ab{f(\varphi)+f''(\varphi)}\ed\varphi.
\end{align}
For a closed, convex curve, the quantity $f(\varphi)+f''(\varphi)$ is the radius of curvature [Eq.~\eqref{eq:ConvexCurvature}], which is nonnegative.  We therefore have
\begin{align}
	P&=\int_0^{2\pi}\br{f(\varphi)+f''(\varphi)}\ed\varphi\\
	&=\int_0^{2\pi}f(\varphi)\ed\varphi.
\end{align}
Under the decomposition \eqref{eq:AntipodalSplit}, the parity-odd piece $C(\varphi)$ does not contribute to the integral, while the parity-even piece contributes
\begin{align}
	P=\int_0^\pi d(\varphi)\ed\varphi.
\end{align}
Thus, the perimeter of a closed, convex curve is the integral of the projected diameter.  When $d(\varphi)$ is constant, we have Barbier's theorem: the perimeter of a constant-width shape is $\pi d$.

\section{Review of plane curve implicitization}
\label{app:Implicitization}

A classical result in algebraic geometry states that every plane curve with rational parameterization can be recast in implicit form as a polynomial equation.  The original curve is composed of one or more of the roots of this polynomial, with additional real roots corresponding to other parameter ranges in the formula defining the original curve.  The set of all such curves forms an algebraic variety.  The conceptually simplest approach to finding this associated algebraic variety is via the method of resultants.  Since many readers may be unfamiliar with this approach, we present a brief sketch here.

Consider a rational curve parameterized by $\sigma$,
\begin{align}
	\label{eq:RationalCurve}
	\pa{x(\sigma),y(\sigma)}=\pa{\frac{p_1(\sigma)}{p_0(\sigma)},\frac{p_2(\sigma)}{p_0(\sigma)}}.
\end{align}
We may regard this curve as the intersection of two surfaces in $\mathbb{R}^3$ given by the two polynomials
\begin{align}
	\label{eq:Surface1}
	f(\sigma)&:=x(\sigma)p_0(\sigma)-p_1(\sigma)=0,\\
	\label{eq:Surface2}
	g(\sigma)&:=y(\sigma)p_0(\sigma)-p_2(\sigma)=0.
\end{align}

Two polynomials $f(x)=a_nx^n+a_{n-1}x^{n-1}+\cdots+a_0$ and $g(x)=b_mx^m+b_{m-1}x^{m-1}+\cdots+b_0$ admit a common root if and only if the determinant of the $(m+n)\times(m+n)$ Sylvester matrix
\begin{align}
	\mathcal{M}(f,g)=
	\begin{bmatrix}
		a_n & a_{n-1} & \cdots & a_0 & 0 & \cdots & 0 \\
		0 & a_n & \cdots & a_1 & a_0 & \cdots & 0 \\
		\vdots & & \ddots & & & \ddots & \\
		0 & \cdots & 0 & a_n & a_{n-1} & \cdots & a_0 \\
		b_m & b_{m-1} & \cdots & b_0 & 0 & \cdots & 0 \\
		0 & b_m & \cdots & b_1 & b_0 & \cdots & 0 \\
		\vdots & & \ddots & & & \ddots & \\
		0 & \cdots & 0 & b_m & b_{m-1} & \cdots & b_0
	\end{bmatrix}
\end{align}
vanishes.  The determinant $\det\mathcal{M}(f,g)$ of this Sylvester matrix is known as the resultant of the two polynomials $f$ and $g$.  Since the condition $\det\mathcal{M}=0$ is satisfied at the intersection of the surfaces (i.e., on the rational curve), but does not involve the parameter $\sigma$, it provides the desired implicit form.

\section{The Kerr critical locus as an algebraic variety}
\label{app:AlgebraicGeometry}

The basic object of study in algebraic geometry is an algebraic variety (the set of solutions to a system of polynomials).  As reviewed in App.~\ref{app:Implicitization}, the implicitization of a rational plane curve always results in a polynomial (the resultant) whose vanishing locus (an algebraic variety) contains the plane curve.  In general, this vanishing locus may also include other curves that are traced by the original parameterization with different ranges of the parameter.  In such cases, rather than examine the original curve individually, it may be more profitable to view it as a subset of its containing algebraic variety.  Here, we adopt this algebro-geometric perspective and study the Kerr critical curve as a subset of its associated algebraic variety, which we dub the \textit{critical locus} to distinguish it from the physical critical curve.  A similar approach was adopted for the study of timelike orbits in Ref.~\cite{Stein2020}.

Following Ref.~\cite{GrallaLupsasca2020b}, we will first study the critical curve in the space of photon conserved quantities $(\lambda,\eta)$, rather than on the image plane $(\alpha,\beta)$.  This is because there is not just one image plane but rather infinitely many, one for each observer inclination $\theta_o$.  By working in $(\lambda,\eta)$-space, we can derive general results that map back to the image plane(s) via Eqs.~\eqref{eq:BardeenCoordinates}.  In this appendix, we will use the inverse form of these equations,
\begin{align}
	\label{eq:ScreenInversion}
	\lambda=-\alpha\sin{\theta_o},\quad
	\eta=\pa{\alpha^2-a^2}\cos{\theta_o}+\beta^2.
\end{align}
For any choice of observer $\theta_o$, each point $(\lambda,\eta)$ defines either two or zero points on the image plane, according to whether there exist real solutions of Eqs.~\eqref{eq:ScreenInversion} for $(\alpha,\beta)$.

We define the $(\lambda,\eta)$-space \textit{critical locus} to be the set of all points $(\lambda(\tilde{r}),\eta(\tilde{r}))$ in the real $(\lambda,\eta)$-plane obtainable from the parameterization \eqref{eq:CriticalParameters} for some real value of $\tilde{r}$.  One may invert $\lambda(\tilde{r})$ to obtain $\tilde{r}(\lambda)$ by solving a cubic equation, resulting in three roots (Eq.~(122) of Ref.~\cite{GrallaLupsasca2020b})
\begin{align}
	\tilde{r}^{(k)}(\lambda)&=M+2M\triangle_\lambda\cos\br{\frac{2\pi k}{3}+\frac{1}{3}\arccos\pa{\frac{1-\frac{a^2}{M^2}}{\triangle_\lambda^3}}},\nonumber\\
	\label{eq:CubicRoots}
	&\triangle_\lambda=\sqrt{1-\frac{a(a+\lambda)}{3M^2}},
\end{align}
with $k\in\cu{0,1,2}$.  In particular, note that Eq.~\eqref{eq:Radius} is recovered as $\tilde{r}^{(0)}(-\alpha\sin{\theta_o})$.  Substituting Eq.~\eqref{eq:CubicRoots} into $\eta(\tilde{r})$ [Eq.~\eqref{eq:CriticalParameters}] defines three functions
\begin{align}
	\label{eq:EtaRoots}
	\eta_i(\lambda):=\eta\pa{\tilde{r}^{(i+1)}(\lambda)},\quad
	i\in\cu{1,2,3},
\end{align}
which together parameterize the entire critical locus.

While this description allows us to write each part of the critical locus as an ordinary function, the price to be paid is that this involves rather complicated irrational functions.  However, returning to the original parameterization \eqref{eq:CriticalParameters} of the critical locus, we see that the latter is a rational curve.  Therefore, as reviewed in App.~\ref{app:Implicitization}, it is possible to recast the critical locus in implicit form as the vanishing locus of a polynomial in $\lambda$ and $\eta$.  Hence, the critical locus is an algebraic variety (though the physical critical curve by itself is not).

From the point of view of the complicated functions $\eta_i(\lambda)$, this is rather remarkable:  Although none has a terminating series expansion in $\lambda$, the \textit{product}
\begin{align}
	\label{eq:RootProduct}
	\mathcal{P}(\lambda,\eta):=\prod_{i=1}^3\br{\eta-\eta_i(\lambda)}
\end{align}
must in fact be a terminating polynomial.  It is by definition cubic in $\eta$, and we shall see that it has degree 6 in $\lambda$.  Furthermore, since the original parameterization \eqref{eq:CriticalParameters} is also rational in $a$, the function $\mathcal{P}$ is a polynomial in $a$ as well.  This fact will be useful for small-spin expansions.

\subsection{Polynomial defining the Kerr critical locus}

We now derive the polynomial defining the Kerr critical locus as an algebraic variety.  Comparing the rational parameterization \eqref{eq:CriticalParameters} with Eqs.~\eqref{eq:RationalCurve}, \eqref{eq:Surface1}, and \eqref{eq:Surface2}, we infer that the associated polynomials $f$ and $g$ may be taken to be
\begin{align}
	f(\tilde{r})&=\tilde{r}^2\pa{\tilde{r}-3M}+a^2\pa{\tilde{r}+M}+a\pa{\tilde{r}-M}\lambda,\\
	g(\tilde{r})&=\tilde{r}^3\br{\tilde{r}\pa{\tilde{r}-3M}^2-4a^2M}+a^2\pa{\tilde{r}-M}^2\eta.
\end{align}
Computing their resolvant yields
\begin{align}
	\det\mathcal{M}\pa{f,g}=4\pa{M^2-a^2}M^2a^6\mathcal{C}(\lambda,\eta),
\end{align}
with
\begin{align}
	\label{eq:CriticalPolynomial}
	\mathcal{C}(\lambda,\eta)&=\pa{M^2-a^2}\eta^3+c_2(\lambda)\eta^2+c_1(\lambda)\eta\nonumber\\
	&\quad+M^2\pa{\lambda-a}^3c_0(\lambda),\\
	c_2(\lambda)&=\pa{3M^2-2a^2}\lambda^2-4a^3\lambda\nonumber\\
	&\quad-\pa{27M^4-33M^2a^2+2a^4},\\
	c_1(\lambda)&=\pa{3M^2-a^2}\lambda^4-4a^3\lambda^3\nonumber\\\
	&\quad-6\pa{9M^4-5M^2a^2+a^4}\lambda^2+4\pa{27M^4-a^4}a\lambda\nonumber\\
	&\quad-\pa{54M^4+33M^2a^2+a^4}a^2,\\
	c_0(\lambda)&=\lambda^3+3a\lambda^2-3\pa{9M^2-a^2}\lambda\nonumber\\
	&\quad+\pa{27M^2+a^2}a.
\end{align}
By construction, the resolvant $\det\mathcal{M}(f,g)$ is a polynomial whose vanishing locus in the real $(\lambda,\eta)$-plane is precisely the Kerr critical locus.  Therefore, it must be proportional to the polynomial $\mathcal{P}$ defined in Eq.~\eqref{eq:RootProduct}.  Comparing the coefficients of their $\O{\eta^3}$ terms then shows that (for $\ab{a}\neq M$)\footnote{When $a=M$, the polynomial $\mathcal{C}$ remains well-defined but degenerates to a simpler quadratic in $\eta$.  Hence it defines the extremal critical locus as a quartic in $\beta$ (see discussion in Sec.~\ref{sec:ExtremalLocus} below).}
\begin{align}
	\mathcal{C}(\lambda,\eta)=\pa{M^2-a^2}\mathcal{P}(\lambda,\eta).
\end{align}
We refer to $\mathcal{C}$ as the \textit{critical polynomial}.

Note that it would have been very challenging (if at all possible) to derive the polynomial expression \eqref{eq:CriticalPolynomial} by explicit multiplication of the product \eqref{eq:RootProduct}.

\subsection{Polynomial description on the image plane}

Now consider the image plane $(\alpha,\beta)$ of an observer with inclination $\theta_o$.  We define the \textit{image plane} critical locus as the $\theta_o$-dependent set of all (real) points $(\alpha,\beta)$ obtained by solving Eqs.~\eqref{eq:ScreenInversion}, when $\lambda$ and $\eta$ lie on the $(\lambda,\eta)$-space critical locus.  This is equivalent to the set of all real curves parameterized by Eqs.~\eqref{eq:BardeenCoordinates} for any real value of the parameter $\tilde{r}$.

As in the previous subsection, we may view this locus either as a set of ordinary functions or as an algebraic variety.  For the former, we use the the local description $\eta_i(\lambda)$ of the $(\lambda,\eta)$-space critical locus \eqref{eq:EtaRoots}.  Plugging into Eq.~\eqref{eq:ScreenInversion} gives a description of the corresponding curves $\beta_i(\alpha)$ (if they exist) on the observer screen.  By the same procedure as in main text [see Eqs.~\eqref{eq:CriticalCurve}--\eqref{eq:F}], they are obtained as the roots of the quadratic equations for $\beta_i$ given by
\begin{align}
	\label{eq:CriticalPieces}
	\mathcal{C}_i:=\alpha^2+\beta_i^2-a^2\cos^2{\theta_o}-F_i(a,\alpha\sin{\theta_o})=0,\\
	F_i(a,\alpha\sin{\theta_o})=\eta_i(-\alpha\sin{\theta_o})+\alpha^2\sin^2{\theta_o},
\end{align}
with $i\in\cu{1,2,3}$.  Note that only a portion of the full critical locus in $(\lambda,\eta)$-space is mapped to the observer screen in this way, namely the portion for which $\beta^2\ge0$.  The actual critical curve is a further subset illustrated in  Fig.~1 of \cite{GrallaLupsasca2020a}.  Note that $F_1$ recovers Eq.~\eqref{eq:F} in the main body, which defines the physical critical curve on the observer screen for sufficiently small spin and inclination.

The situation here is analogous to that described for $(\lambda,\eta)$-space above Eq.~\eqref{eq:RootProduct}:  While explicit, the functions $\mathcal{C}_i(a,\alpha\sin{\theta_o})$ are rather complicated due to their dependence on the cubic roots \eqref{eq:CubicRoots}.  However, the curves $\mathcal{C}_i=0$ together form the vanishing locus of the critical polynomial under the substitution \eqref{eq:ScreenInversion},
\begin{align}
	\label{eq:CriticalPolynomialImage}
	\mathcal{C}_o(\alpha,\beta):=\mathcal{C}\pa{-\alpha\sin{\theta_o},\pa{\alpha^2-a^2}\cos^2{\theta_o}+\beta^2}.
\end{align}
Since $\mathcal{C}(\lambda,\eta)$ is a polynomial in $\lambda$, $\eta$, and $a$, we see from the form of Eq.~\eqref{eq:CriticalPolynomialImage} that $\mathcal{C}_o$ is polynomial in all of $\alpha$, $\beta$, $\sin\theta_o$ (degree 6), and $a$ (degree 8).  More explicitly, the product $\mathcal{C}_1\mathcal{C}_2\mathcal{C}_3$ must be a polynomial proportional to $\mathcal{C}_o$, and comparison of the coefficients of their $\O{\beta^6}$ terms shows that (for $\ab{a}\neq M$),
\begin{align}
	\label{eq:Factorization}
	\mathcal{C}_o(\alpha,\beta)=\pa{M^2-a^2}\mathcal{C}_1\mathcal{C}_2\mathcal{C}_3.
\end{align}
Though each of the expressions $\mathcal{C}_i$ for $i\in\cu{1,2,3}$ is not individually a polynomial (with their expansion in small parameters $a$ or $\sin \theta_o$ producing infinite power series), their product $\mathcal{C}_1\mathcal{C}_2\mathcal{C}_3$ is a polynomial of finite degree.

\subsection{Extremal spin factorization}
\label{sec:ExtremalLocus}

As discussed in Sec.~\ref{sec:ExtremalCriticalCurve}, the extremal limit $a\to M$ of the critical curve is rather subtle, requiring two separate $a\to M$ limits \cite{Gralla2018}.  Here we have passed to the critical \textit{locus} and described it as an algebraic variety, which depends only on the physical parameters $a$ and $M$, and hence has only a single limit $a\to M$.  Setting $a=M$ in Eq.~\eqref{eq:CriticalPolynomial}, the critical polynomial $\mathcal{C}(\lambda,\eta)$ degenerates to a quadratic polynomial in $\eta$, which factorizes into
\begin{align}
	\mathcal{C}(\lambda,\eta)&=M^2\pa{\lambda-2M}^2\mathcal{E}(\lambda,\eta),\\
	\mathcal{E}(\lambda,\eta)&=\eta^2+2\pa{\lambda^2+2M\lambda-11M^2}\eta\nonumber\\
	&\quad+\pa{\lambda-M}^3\pa{\lambda+7M}.
\end{align}
Upon mapping to the image plane $(\alpha,\beta)$ via Eq.~\eqref{eq:ScreenInversion}, the (multiplicity 2) factor $\lambda-2M=0$ defines a double line $\alpha=-2M\csc{\theta_o}$, a portion of which forms the NHEKline that becomes visible when $\sin{\theta_o}>\sqrt{3}-1$.  On the other hand, the remaining factor becomes a quartic polynomial in both $\alpha$ and $\beta$, already given as Eq.~\eqref{eq:CartesianOvals} in the main text, which defines a pair of Cartesian ovals.  The full extremal critical locus is illustrated in Fig.~\ref{fig:ExtremalCurves}.  Note that the infinite vertical line is associated with roots of $\mathcal{C}_o$ that become complex for any $a<M$; no vestige of this feature remains for non-extremal black holes.

\subsection{Small spin factorization}

In the case of a non-rotating Schwarzschild black hole ($a=0$), the critical polynomial $\mathcal{C}_o(\alpha,\beta)$ factorizes as
\begin{align}
	\mathcal{C}_o(\alpha,\beta)=M^2\pa{\alpha^2+\beta^2}^2\pa{\alpha^2+\beta^2-27M^2}.
\end{align}
The vanishing locus of this polynomial is the circular critical curve of radius $3\sqrt{3}M$.  Note the ``circle of zero radius'' factor $\alpha^2+\beta^2$.  At higher order in spin, this becomes a second piece of the critical locus, a small oval inside the larger critical curve.  We have not factorized the polynomial at higher order in spin.  However, we have checked that the polynomial vanishes to $\O{a^2}$ when evaluated on the ellipse ansatz \eqref{eq:CanonicalEllipse} with the given parameters \eqref{eq:CenterOffset}, \eqref{eq:RadiusCorrection1}, and \eqref{eq:RadiusCorrection2}.

\subsection{Small inclination factorization}

The image plane of a polar observer ($\theta_o=0$) is best described in terms of the impact parameter $b=\sqrt{\alpha^2+\beta^2}$ (see Eq.~(61) of Ref.~\cite{GrallaLupsasca2020a}).  On the pole, the critical polynomial $\mathcal{C}_o(\alpha,\beta)$ defined in Eq.~\eqref{eq:CriticalPolynomialImage} reduces to a cubic polynomial in $b^2$,
\begin{align}
	\mathcal{C}_o(b)&=\pa{M^2-a^2}b^6-\pa{27M^4-30M^2a^2-a^4}b^4\nonumber\\
	\label{eq:CriticalPolynomialPole}
	&\quad-96M^2a^4b^2+64M^2a^6=0.
\end{align}
Its 6 roots (4 are real and 2 complex conjugates) are
\begin{align}
	b=\pm\sqrt{\eta_i(0)+a^2},\quad
	i\in\cu{1,2,3},
\end{align}
with $\eta_i(\lambda)$ as defined in Eq.~\eqref{eq:EtaRoots}.  Two of these roots satisfy $b>0$, and hence are part of the image plane critical locus.  The actual critical curve has radius (see Eq.~(67) of Ref.~\cite{GrallaLupsasca2020a})
\begin{align}
	\tilde{b}=\sqrt{\eta_1(0)+a^2}.
\end{align}

This formula allows us to pick out the physical part of the critical locus (i.e., the critical curve) at higher order in $\sin{\theta_o}$.  Using the ellipse ansatz \eqref{eq:CanonicalEllipse},
\begin{align}
	\beta^2=R_2\br{1-\pa{\frac{\alpha-\alpha_0}{R_1}}^2},
\end{align}
together with $R_1=\tilde{b}+\O{\sin\theta_o}$ and $R_2=\tilde{b}+\O{\sin\theta_o}$, we plug into the critical polynomial \eqref{eq:CriticalPolynomialImage} and find that to leading order $\O{\sin{\theta_o}}$,
\begin{align}
	\label{eq:SmallInclination}
	\alpha_0&=\tilde{b}^2Xa\sin{\theta},\quad
	R_1=R_2=\tilde{b},
\end{align}
where
\begin{align}
	\label{eq:X}
	X&=-\frac{2\pa{a^2\tilde{b}^2-27M^4-a^4}}{3\pa{M^2-a^2}\tilde{b}^4-2\mathcal{X}\tilde{b}^2-96M^2a^4},\\
	\mathcal{X}&=27M^4-30M^2a^2-a^4.
\end{align}
Plotting numerically shows that $\alpha_0>0$ for all positive spins $0<a\le M$.  That is, the critical curve remains circular at linear order in inclination, with a rightward origin shift given by Eqs.~\eqref{eq:SmallInclination} and \eqref{eq:X}.

Repeating the procedure at next order $\O{\sin^2{\theta_o}}$, we find the ellipse parameters
\begin{align}
	\label{eq:ParametersStart}
	\alpha_0&=\tilde{b}^2Xa\sin{\theta},\\
	R_1&=\tilde{b}-\frac{X^3Y\pa{a\sin{\theta}}^2}{16\tilde{b}\pa{a^2\tilde{b}^2-27M^4-a^4}^3\pa{M^2-a^2}^3},\\
	R_2&=\tilde{b}+\frac{\tilde{b}^4X^2-1}{2\tilde{b}}\pa{a\sin{\theta}}^2,
\end{align}
where
\begin{align}
	Y&=64M^4a^8Z_0+32\pa{9M^2-7a^2}M^4a^6Z_2\tilde{b}^2\nonumber\\
	&\quad-M^2a^2Z_4\tilde{b}^4,\\
	Z_0&=31a^{12}-1674M^2a^{10}+7053M^4a^8-12636M^6a^6\nonumber\\
	&\quad+12393M^8a^4-7290M^{10}a^2+2187M^{12},\\
	Z_2&=25a^{10}-721M^2a^8+2382M^4a^6-3078M^6a^4\nonumber\\
	&\quad+2025M^8a^2-729M^{10},\\
	Z_4&=a^{16}-3544M^2a^{14}+61796M^4a^{12}\nonumber\\
	&\quad-309528M^6a^{10}+746334M^8a^8\nonumber\\
	&\quad-1006344M^{10}a^6+784404M^{12}a^4\nonumber\\
	\label{eq:ParametersEnd}
	&\quad-332424M^{14}a^2+59049M^{16}.
\end{align}
To produce these equations, we have used Eq.~\eqref{eq:CriticalPolynomialPole}, which is obeyed by $\tilde{b}$, to reduce the degree of the polynomials in $\tilde{b}$.  The initial result of solving for $R_1$ yielded a polynomial of degree 14 in the irrational expression $\tilde{b}$. 

\bibliography{reconstruct}
\bibliographystyle{utphys}

\end{document}